\documentclass[aps,
               prc,
               twocolumn,
               superscriptaddress,
               floatfix,
               amsmath,
               amssymb,
               nofootinbib,
               longbibliography]{revtex4-1}
\usepackage{bm}
\usepackage{svg}
\usepackage{braket}
\usepackage{amsmath}
\usepackage{amssymb}
\usepackage{color}
\usepackage{epsfig}
\usepackage{MnSymbol}
\usepackage{verbatim}
\usepackage{wrapfig}
\usepackage{pifont}
\usepackage{amsfonts}
\usepackage{slashed}
\usepackage{latexsym}
\usepackage{psfrag}
\usepackage{nicefrac}
\usepackage{ragged2e}
\usepackage{etoolbox}
\usepackage{tikz}
\usepackage{pgfpages}
\usepackage{dsfont}
\usepackage{graphics}
\usepackage{graphicx}
\usepackage{comment}

\usepackage[ocgcolorlinks,citecolor=blue,linkcolor=blue,urlcolor=blue]{hyperref}

\begin{document}

\title {Removing center of mass effects in response function and sum rule calculations based on the harmonic oscillator basis}

\author{K.~Akdogan} 
\affiliation{Institut f\"ur Kernphysik and PRISMA$^+$ Cluster of Excellence, Johannes Gutenberg-Universit\"at, 55128
  Mainz, Germany}

\author{D.~Layh}
\affiliation{Institut f\"ur Kernphysik and PRISMA$^+$ Cluster of Excellence, Johannes Gutenberg-Universit\"at, 55128
  Mainz, Germany}

\author{I.~Weinberger}\affiliation{Racah Institute of Physics, Hebrew
  University, 91904, Jerusalem}

\author{J.~Simonis}
\affiliation{Institut f\"ur Kernphysik and PRISMA$^+$ Cluster of Excellence, Johannes Gutenberg-Universit\"at, 55128
  Mainz, Germany}

\author{N.~Barnea} \affiliation{Racah Institute of Physics, Hebrew
  University, 91904, Jerusalem}
  
\author{S.~Bacca}
\affiliation{Institut f\"ur Kernphysik and PRISMA$^+$ Cluster of Excellence, Johannes Gutenberg-Universit\"at, 55128
  Mainz, Germany}
  \affiliation{Helmholtz-Institut Mainz, Johannes Gutenberg-Universit\"at Mainz, D-55099 Mainz, Germany}

\vspace{1cm}
\date{\today}

\begin{abstract}
Response functions are at the heart of any comparison of theory with experiment in studies of the nuclear dynamics  with electroweak probes. Calculations performed in the laboratory frame often suffer from  center of mass contaminations that need to be removed. By confining the system in a harmonic oscillator, we derive a set of analytical formulas to subtract the  center of mass effects from calculations of
response functions and associated sum rules. After  a general analytical derivation, we first deal specifically with the longitudinal response function appearing in electron scattering and provide expressions for the center of mass correcting functions. Next, we present a proof of principle study for the case of the electric dipole sum rules in a two-body problem with a numerical implementation of our formalism. These steps pave the way to applying the proposed method to heavier nuclei  in the future.
\end{abstract}
\pacs{find pacs}

\maketitle

%%%%%%%%%%%%%%%%%%%%%%%%%%%%%%%%%%%%%% SECTION %%%%%%%%%%%%%%%%%%%%%%%%%%%%%%%%%%%%%%

\section{introduction}
The nucleus is a self bound system, made by many interacting protons and neutrons. In contrast to an atomic system, when solving the nucleus as a quantum many-body problem one does not have an external potential that provides confinement nor  a natural center for the coordinate system. Hence, one typically refers the $A$-nucleons
 coordinates $\left\{{\bf r}_k \right\}$  to an arbitrarily fixed coordinate center. One possible choice for the coordinate center is  the laboratory frame, another possible choice is the center of mass (CoM) frame,  also called internal or intrinsic frame.
 After defining the CoM coordinate for $A$ equal-mass particles as
\begin{equation}
 {\bf R}_{\rm CoM} = \frac{1}{A}\sum_k {\bf r}_k  \,,   
\end{equation}
 one may, in fact, introduce the internal coordinates  
\begin{equation}
\label{rel}
 {\bf r'}_k={\bf r}_k - {\bf R}_{\rm CoM}\,,
\end{equation}
where one essentially takes the CoM as the reference center of the coordinate system.

%The latter are not linearly independent, because $\sum_k {\bf r'}_k =0$.

In terms of the single-particle coordinates $\left\{{\bf r}_k \right\}$, 
the nuclear Hamiltonian is
the sum of the kinetic and potential energy 
\begin{eqnarray}
\label{start}
\hat{\mathcal H}  & = & \hat{T}+\hat{V} = \\
\nonumber
 & =& \sum_k^A \frac{{\bf p}_k^2}{2m_N} + \sum_{k<j}^A \hat{V}({\bf r}_k - {\bf r}_j)\,,
\end{eqnarray}
where $m_N$  denotes the nucleon mass, an average of the proton and neutron masses, and  where only two-body forces are considered without writing explicitly their spin-isospin dependence\footnote{Note that
adding three-body forces does not change the presented formalism.}. 
The potential $\hat{V}$ in Eq.~(\ref{start}) depends only on the difference of the particle-coordinates and is thus translational invariant, namely it is the same independently on the center of the reference frame.
%whether one uses $\left\{{\bf r}_k \right\}$ or $\left\{{\bf r}'_k \right\}$ as coordinates.

%Internal coordinates are the only coordinates needed to describe the dynamics of a nucleus.

The Hamiltonian (\ref{start}) can be written as
\begin{equation}
\label{H_ansatz}
\hat{\mathcal H}   =  \hat{H}^{\rm I}+ \hat{T}^{\rm CoM}\,, 
\end{equation}
where $\hat{T}^{\rm CoM} = \frac{{\bf P}^2_{\rm CoM}}{2m_NA}$
 is the CoM kinetic energy 
 and $\hat{H}^{\rm I}$ is the internal (I) Hamiltonian 
\begin{eqnarray}
\label{int}
\hat{H}^{\rm I} &=& \sum_{k<j}^A \frac{({\bf p}_k - {\bf p}_j)^2}{2Am_N} + \sum_{k<j}^A \hat{V}({\bf r}_k - {\bf r}_j) =\\
\nonumber
&=& \sum_{k<j}^A \frac{({\bf p}'_k - {\bf p}'_j)^2}{2Am_N} + \sum_{k<j}^A \hat{V}({\bf r}'_k - {\bf r}'_j)  \,,
\end{eqnarray}
where ${\bf p}_k'$ are the momenta relative to the CoM momentum.
It is to note that $\hat{H}^{\rm I}$ is translational invariant, namely it has the exact same form, e.g., in terms of particle-coordinates $\left\{{\bf r}_k \right\}$ referred to the laboratory frame or in term of the internal coordinate
$\left\{{\bf r}'_k \right\}$ referred to the CoM.
%Here and thereafter, we will call the first choice as the laboratory frame and the second as the intrinsic frame coordinate.

Because relative coordinates (\ref{rel}) are not linearly independent, it is customary to transform 
 the $A$-particles coordinate system $\left\{{\bf r}_k \right\}$  into a system where coordinates are made by ${\bf R}_{\rm CoM}$
and $(A-1)$ linearly independent relative position vectors, called the Jacobi coordinates~\cite{Jacobi}.  
The latter can be used to describe the internal dynamics of $\hat{H}^{\rm I}$.
%Working with Jacobi coordinates or the relative coordinates (\ref{rel})
%is equivalent as setting the center of the coordinate system exactly in the CoM.
Working with Jacobi coordinates and dealing only with $\hat{H}^{\rm I}$, disregarding completely the CoM, is the preferred choice when studying light nuclei (see, e.g., Refs.~\cite{Navratil2000,barnea2001}). Solving for $\hat{H}^{\rm I}$ is indeed the only interesting part in the description of nuclear properties.
However, the antisymmetrization of the wave function is typically the limiting factor in extending this approach to nuclei with $A>7$~\cite{Bacca:2009yk,Bacca:2013ayo}. To study medium  to heavy  nuclei, the typical choice is that of the single-particle coordinates $\left\{{\bf r}_k \right\}$, because antisymmetrization is easily obtained with Slater determinants defined in terms of single-particle states~\cite{Slater1929}.

Clearly, whenever the Hamiltonian can be written as the sum of an intrinsic  and a CoM component as in Eq.~(\ref{H_ansatz}), its eigenstates can be factorized into a product of the CoM wave function and the internal wave function
\begin{equation}
\label{fact_wf}
\ket{\Psi} =   \ket{\Psi^{\rm I}}\otimes \ket{\Psi}^{\rm CoM} 
\end{equation}
with the corresponding energy given by the sum of the two subsystem energies ${\mathcal E}=\varepsilon+E$. Here, we use the Greek letter for the internal energy and the Latin letter for the CoM energy.
If the Hamiltonian can be written as in Eq.~(\ref{H_ansatz}),   the separation of the CoM component from the internal part is ensured by
(\ref{fact_wf}), so that  knowing $\ket{\Psi}^{\rm CoM}$ one can 
extract the internal wave functions, even when working with a non-translational invariant Hamiltonian. 
Whether the factorization in Eq.(\ref{fact_wf}) is preserved or not,  depends on the method used to represent the nuclear wave function and to solve the Hamiltonian problem.
%where the CoM wave function is a solution of the free Schr\"odinger equation given by plane waves.

One way to solve the Schr\"odinger equation for the nuclear Hamiltonian  is to expand the wave function on a complete set of basis states. After truncation of that basis, one turns to the task of diagonalizing the Hamiltonian matrix on the basis of choice. The favorite basis in nuclear physics is the harmonic oscillator basis, the traditional application being the nuclear shell model~\cite{deShalit}.
With the harmonic oscillator basis defined in terms of single-particle coordinates $\left\{{\bf r}_k \right\}$, the exact factorization of Eq.~(\ref{fact_wf}) is preserved only when one uses the harmonic oscillator in a complete $N \hbar \Omega$ space~\cite{Talmi}, where
$N$ is the harmonic oscillator quantum number and $\hbar\Omega$ the harmonic oscillator frequency. When the basis is truncated in any other way, the factorization may not hold.
This causes the appearance of "spurious CoM states" which contaminate the theoretical description of the internal dynamics.
%when one uses a Hamiltonian which is not just the internal $H^{\rm I}$. 
This problem, also known as "CoM problem",  has been discussed in the literature very early in the history of nuclear many-body  theory~\cite{gartenhaus1957,Talmi,gloeckner1974,mcgrory1975}, but also more recently, where various solutions have been proposed (mostly for ground-state properties)  for no-core shell model~\cite{roth2009b}
or coupled-cluster theory calculations~\cite{mihaila1999,hagen2009a}. 

In this paper, we want to analyze the case of response functions and related sum rules, which involve the excitation of the nucleus. Recently, it was shown that the Lawson method~\cite{gloeckner1974} can be numerically implemented to treat the CoM spurious states in sum rule calculations~\cite{Baker2020}.  Here, we want instead to
present an analytical method to remove center
of mass effects in the case of response function calculated in single particle basis. 
The method is valid for any such basis, as long as the ground state and response functions are calculated with high enough accuracy.
Nevertheless, we expect it to work best for the harmonic oscillator basis.

When working with the harmonic oscillator basis, one typically adds  a CoM  potential $\hat{V}^{\rm CoM}=\frac{1}{2}Am_N \Omega {\bf R}^2_{\rm CoM}$ to $\hat{\mathcal H}$, so that the new Hamiltonian becomes
\begin{equation}
\label{H_ansatz2}
 \hat{H}   =  \hat{H}^{\rm I}+ \hat{H}^{\rm CoM}\,, 
\end{equation}
with $\hat{H}^{\rm CoM} = \hat{T}^{\rm CoM}+ \hat{V}^{\rm CoM}$.
This Hamiltonian obviously admits the factorization of Eq.~(\ref{fact_wf}) and the CoM wave functions $\ket{\Psi}^{\rm CoM}$
are known harmonic oscillator eigenstates.

In the following, we will assume that $\hat{H}$ has already been solved in the coordinates $\left\{{\bf r}_k \right\}$, which we call the laboratory frame, and we will derive an analytical connection to calculations in the internal frame governed by $\hat{H}^{\rm I}$. 
Our goal is to derive such connections for response functions and sum rules   that are induced by
external operators which can be written in the following form
\begin{equation}\label{eq:Operator decoupling}
\hat{O} = \hat{O}^{\rm I} \, \hat{O}^{\rm CoM},
\end{equation}
with $\hat{O}^{\rm I}$ acting only on the internal wave function,
and $\hat{O}^{\rm CoM}$ on the CoM one.
Most of the operators used in electroweak processes have this form~\cite{th4He,bijaya} and we will show later that our formalism can be generalized also to the case where $\hat{O} = \hat{O}^{\rm I} + \hat{O}^{\rm CoM}$, which include another large category of electroweak operators.

The paper is structured as follows. 
In Section~\ref{section:resp_func} and Section~\ref{section:sum_rule}
we present the analytical derivation of our method for response functions and related sum rules, respectively. In Section~\ref{section:el_scatt} we present first the example of the electron scattering process, while in Section~\ref{section:photoabs}
we deal with the photoabsorption case as a second example, where we provide a numerical implementation of our analytical formulae. Finally, in Section~\ref{section:conclusions} we draw our conclusions.

%%%%%%%%%%%%%%%%%%%%%%%%%%%%%%%%%%%%%% SECTION %%%%%%%%%%%%%%%%%%%%%%%%%%%%%%%%%%%%%%

\section{The response function}
\label{section:resp_func}

 The response function is the basic  quantity that describes the interaction of a nuclear system with external probes and is defined as~\cite{Bacca_2014}
\begin{equation}\label{rf}
R (\omega) =  \sum_{F,\bar{0}}
\left| \braket{\Psi_{F}|\hat{O}|\Psi_{0}}\right|^{2}
\delta \left({\mathcal E}_F-{\mathcal E}_0-\hbar\omega\right) \,.
\end{equation}
Here, $\omega$ is the energy transfer, $\Psi_0$ and $\Psi_F$ are the
initial (ground state) and final state wave functions with energy ${\mathcal E}_{0/F}$, respectively, while $\hat{O}$ is the excitation operator, determined by the probe. The summation symbol in Eq.~(\ref{rf}) has to be understood as a sum on all discrete and an integral on all the continuous quantum numbers in the final state, cumulatively denoted by $F$. An average on the initial state quantum numbers is also typically included in the definition and is  denoted here with $\bar{0}$. Finally, the delta function ensures the conservation of the energy.

Let us  assume that we have already solved this problem in the laboratory frame working with coordinates $\left\{{\bf r}_k \right\}$.
Our goal is to derive a formalism to separate the total response function of Eq.~(\ref{rf}) calculated in the laboratory frame into  a center of mass term ---to be removed---and the intrinsic response function, that pertains to the degrees of freedom included in $\hat{H}^{\rm I}$ and $\hat{O}^{\rm I}$. The latter is what one would directly obtain working  in the center of mass frame, e.g., using Jacobi coordinates~\cite{gazit2006,BaccaPRL2009,BaccaMono,Barrett:2013nh,Navratil:2009ut}.
 Because
\begin{equation}\label{eq:indipendent Hamiltonians}
\left[\hat{H}^{\rm I},\hat{H}^{\rm CoM}\right]=0\,,
\end{equation}
we can take the eigenstates of $\hat{H}^{\rm I}$, which we call
$| \Psi_{\varepsilon j m} \rangle$, and those of $\hat{H}^{\rm CoM}$,
which we call $|\Psi_{EJM}^{\rm CoM}\rangle$, and choose
the coupled outer product~\cite{Varshalovich}
\begin{equation}\label{eq:w.f. decoupling}
\ket{\Psi_{{\mathcal E} {\mathcal J}{\mathcal M}}} = \left[ \ket{\Psi_{\varepsilon j}^{\rm I}\rangle\otimes|\Psi_{EJ}^{\rm CoM}}\right]^{\mathcal J}_{\mathcal M} \, 
\end{equation}
as basis for $\hat{H}$. This basis
 simultaneously diagonalizes $\hat{H}^{\rm I}$, $\hat{H}^{\rm CoM}$  and $\hat{H}$ and constitutes a set of states with good angular momentum quantum
numbers $j$ (intrinsic), $J$ (CoM) and ${\mathcal J}$ (total)
with  angular momentum projection  $m$ (intrinsic), $M$ (CoM) and ${\mathcal M}$ (total). 

For the initial CoM state we have $J_0=M_0=0$ so that the initial state becomes
\begin{equation}\label{eq:Initial state decoupling}
\ket{\Psi_0} \rightarrow \ket{\Psi_{{\mathcal E}_0{\mathcal J}_0 {\mathcal M}_0}} = \ket{\Psi_{\varepsilon_0 j_0 m_0}^{\rm I}} \ket{\Psi_{E_000}^{\rm CoM}}, 
\end{equation}
with ${\mathcal E}_0=\varepsilon_0+E_{0}$. For the final state, in Eq.~(\ref{eq:w.f. decoupling}) we write the coupling explicitly
using Clebsch-Gordan coefficients \cite{Varshalovich}, so that the final state in our basis becomes
\begin{align}
\label{eq:Final state decoupling}
\ket{\Psi_F} \rightarrow \ket{\Psi_{{\mathcal E}{\mathcal J}{\mathcal M}}} &= \ket{\Psi_{{\mathcal E}(jJ){\mathcal J}{\mathcal M}}} =\\
 \nonumber
&= \sum_{mM} \braket{jmJM| {\mathcal J}{\mathcal M}} \ket{\Psi_{\varepsilon jm}^{\rm I}} \ket{\Psi_{EJM}^{\rm CoM}} 
\end{align}
with  ${\mathcal E}=\varepsilon +E$. We note that due to rotational invariance the energies do not depend on the angular momentum projection.

In order to calculate the response function of Eq.~({\ref{rf}}) we need to consider   
the transition amplitude 
$\braket{\Psi_{{\mathcal E}{\mathcal J}{\mathcal M}}|\hat{O}|\Psi_{{\mathcal E}_{0}{\mathcal J}_{0}{\mathcal M}_{0}}}$
from an initial state $|\Psi_{{\mathcal E}_{0}{\mathcal J}_{0}{\mathcal M}_{0}}\rangle$ to a final state $|\Psi_{{\mathcal E}{\mathcal J}{\mathcal M}}\rangle$, take the square of it and perform a
 sum over all possible final states and average on the initial state quantum numbers.  
In our notation, the total response function of Eq.~(\ref{rf}) becomes
\begin{eqnarray}\label{seven}
R (\omega) &=& \frac{1}{2 \mathcal{J}_0 + 1} \sum_{\mathcal{M}_0} \sum_{\mathcal E}  \\
\nonumber
&&\sum_{{\mathcal J}{\mathcal M}} \sum_{jJ}
\left| \braket{\Psi_{{\mathcal E}(jJ){\mathcal J}{\mathcal M}}|\hat{O}|\Psi_{{\mathcal E}_0{\mathcal J}_0{\mathcal M}_0}}\right|^{2}
\!\!\delta \left({\mathcal E}-{\mathcal E}_0-\hbar\omega\right), 
\end{eqnarray}
%\begin{widetext}
% equation for widetext
%\begin{equation}\label{seven}
%R (\omega) = \frac{1}{2 \mathcal{J}_0 + 1} %\sum_{\mathcal{M}_0} \sum_{{\mathcal E}{\mathcal %J}{\mathcal M}} \sum_{jJ}
%\left| \braket{\Psi_{{\mathcal E}(jJ){\mathcal %J}{\mathcal M}}|\hat{O}|\Psi_{{\mathcal %E}_0{\mathcal J}_0{\mathcal M}_0}}\right|^{2}
%\delta \left({\mathcal E}-{\mathcal %E}_0-\hbar\omega\right) \,,
%\end{equation}
%where $\omega$ is the energy transferred.
where we note that the $\sum_{\mathcal J, j, J}$ are not free sums, but are connected to each other by angular momentum coupling rules.
The summation $\sum_{\mathcal{E}}$ has to be intended as the sum of any quantum number the energy may depend on, but the angular momentum.

Using Eqs.(\ref{eq:Initial state decoupling}) and (\ref{eq:Final state decoupling}), the squared matrix element in Eq.~(\ref{seven}) becomes
\begin{eqnarray}
\label{sq_me}
&&\sum_{{\mathcal J}{\mathcal M}} \sum_{jJ}
\left| \braket{\Psi_{{\mathcal E}(jJ){\mathcal J}{\mathcal M}}|\hat{O}|\Psi_{{\mathcal E}_0{\mathcal J}_0{\mathcal M}_0}}\right|^{2}=\sum_{{\mathcal J}{\mathcal M}}\sum_{jJ}  \times \\
\nonumber
&& \left|\sum_{mM}\!\! \braket{jmJM|{\mathcal J}{\mathcal M}}\!\!
\bra{\Psi_{\varepsilon jm}^{\rm I}} \braket{\Psi_{EJM}^{\rm CoM}|\hat{O}|\Psi_{\varepsilon_0 j_0m_0}^{\rm I}}
\!\ket{\Psi_{E_000}^{\rm CoM}} \right|^{2}\!\!\!= \\
\nonumber
&&\!\! \sum_{{\mathcal J}{\mathcal M}} \!\!\sum_{jJ} \!\sum_{mM}\!\! \!\braket{jmJM\vert {\mathcal J}{\mathcal M}}\!\!
\bra{\Psi_{\varepsilon jm}^{\rm I}}\!\! \braket{\Psi_{EJM}^{\rm CoM}|\hat{O}|\Psi_{\varepsilon_0 j_0m_0}^{\rm I}}\!\!
\ket{\Psi_{E_000}^{\rm CoM}}\!\times  \\
\nonumber
&&\! \left[\! \sum_{m'M'} \!\! \braket{jm'JM' | {\mathcal J}{\mathcal M}}\!\! \bra{\Psi_{\varepsilon jm'}^{\rm I}}\!\!
\braket{\Psi_{EJM'}^{\rm CoM}|\hat{O}|\Psi_{\varepsilon_0 j_0m_0}^{\rm I}}\!\! \ket{\Psi_{E_000}^{\rm CoM}}\!\! \right]^{\star}.
\end{eqnarray}
Because of the orthogonality of the Clebsch-Gordon coefficients
 the squared matrix element of Eq.~(\ref{sq_me}) becomes~\cite{edmonds1996angular}
\begin{eqnarray}
 &&\sum_{{\mathcal J}{\mathcal M}} \sum_{jJ}
\left| \braket{\Psi_{{\mathcal E}(jJ){\mathcal J}{\mathcal M}}|\hat{O}|\Psi_{{\mathcal E}_0{\mathcal J}_0{\mathcal M}_0}}\right|^{2}=\\
\nonumber
&& \underset{jJ}{\sum}\underset{mM}{\sum}\langle\Psi_{\varepsilon jm}^{\rm I}|\langle\Psi_{EJM}^{\rm CoM}|\hat{O}|\Psi_{\varepsilon_0  j_0m_0}^{\rm I}\rangle|\Psi_{E_000}^{\rm CoM}\rangle\\
&&\nonumber \left[\langle\Psi_{\varepsilon jm}^{\rm I}|\langle\Psi_{EJM}^{\rm CoM}|\hat{O}|\Psi_{\varepsilon_0 j_0m_0}^{\rm I}\rangle|\Psi_{E_000}^{\rm CoM}\rangle\right]^{\star}\,.
\end{eqnarray}

Finally, using the factorization ansatz of Eq.~(\ref{eq:Operator decoupling}),  we obtain the
following expression for the response function
\begin{eqnarray}
\label{eq:Separated_response}
R (\omega)  &=& \frac{1}{2 j_0 + 1} \sum_{m_0} \sum_{\mathcal{E}} \sum_{jm}
\left| \braket{ \Psi_{\varepsilon jm}^{\rm I} |  \hat{O}^{\rm I} | \Psi_{\varepsilon_0 j_0 m_0}^{\rm I}} \right|^{2} \\
\nonumber
&&\sum_{JM} \left| \braket{\Psi_{EJM}^{\rm CoM}|\hat{O}^{\rm CoM}|\Psi_{E_000}^{\rm CoM}} \right|^{2}
\delta \left({\mathcal E} - {\mathcal E}_0 - \hbar \omega \right) \,.
\end{eqnarray}
It is to note that due to the presence of the delta function that depends on both the intrinsic
and the CoM energies, we can not use completeness and perform the sum over the CoM quantum numbers $J,M$. 

%At this point we recall that we want to calculate the intrinsic response function $R^{\rm I}\left(\omega\right)$ and that
%Eq.~(\ref{eq:Separated_response}) is the response function of the whole system. 

Next, one realizes that given  $\hat{O}^{\rm CoM}$ it is possible to calculate the CoM
matrix elements $\langle\Psi_{EJM}^{\rm CoM}|\hat{O}^{\rm CoM}|\Psi_{E_000}^{\rm CoM}\rangle$.
Indeed, they can even be
calculated analytically most of the time, since the CoM wave functions are  harmonic oscillator states.
%For coupled-cluster theory, Ref.~\cite{Hagen} showed that wave functions factorize into CoM and intrinsic parts and that the CoM 
%components are eigenstates of an harmonic oscillator corresponding to a frequency $\tilde{\omega}$, which in general is different from 
%the frequency of the underlying single particle basis.  Ref.~\cite{Hagen} also indicates a prescription of how to calculate such an  $
%\Omega$.
%This is the case in a no-core shell model expansion (CITE), but it has been shown to be the case in coupled-cluster theory calculations as well~\cite{hagen2009a}.
%Then, our strategy will be to find the connection between the total response in Eq.~(\ref{eq:Separated_response}) and the intrinsic
%response $R^{\rm I}\left(\omega\right)$ and derive a prescription to recover the latter from the first.
Furthermore, one can rewrite the delta function as
\begin{displaymath}
\delta \left({\mathcal E}-{\mathcal E}_0-\hbar\omega\right) = \delta \left(E-E_0 + \varepsilon - \varepsilon_0-\hbar \omega \right) 
\end{displaymath}
and because the CoM is in a harmonic oscillator state,   excitation energies  will be quantized so as to satisfy
\begin{equation}\label{eq:COM quantization}
 E - E_0 = N \hbar \Omega \; , \qquad N = 0,1,2, \ldots \, ,
\end{equation}
with $N=2 N_r + J$, where $N_r$ is the radial quantum number and $J \leq N$.
This allows one to rewrite Eq.~(\ref{eq:Separated_response}) as
%\begin{widetext}
\begin{equation}\label{eq:respne_total1-2}
\begin{split}
R (\omega) &= \frac{1}{2 j_0 + 1} \sum_{m_0} \sum_{\varepsilon jm} \left| \braket{\Psi_{\varepsilon jm}^{\rm I}|\hat{O}^{\rm I}|\Psi_{\varepsilon_0 j_0m_0}^{\rm I}} \right|^{2} \\
&\times \sum_{NJM} \left| \braket{\Psi_{NJM}^{\rm CoM}|\hat{O}^{\rm CoM}|\Psi_{N_000}^{\rm CoM}} \right|^{2} \\
&\times \delta \left(\varepsilon - \varepsilon_0 + N \hbar \Omega - \hbar \omega \right) \,,
\end{split}
\end{equation}
%\end{widetext}
where $\sum_{\mathcal E}$ has been split into a sum on the internal energy  $\sum_{\varepsilon} $ and a sum on the CoM energy. Because the latter only depends on 
the quantum number ${N}$, the sum on the energy becomes  $\sum_N$ and the label $E$ in the wave function can be substituted just with $N$ as $\ket{\Psi_{EJM}^{\rm CoM}} \rightarrow \ket{\Psi_{NJM}^{\rm CoM}}$, with $N=0$ for the ground state.

Next, we define the CoM transition probabilities as
\begin{equation}\label{transitionprob}
K_N^{\rm CoM} \equiv \sum_{J\leq N} \sum_{M=-J}^{+J} \left| \braket{\Psi_{N JM}^{\rm CoM}|\hat{O}^{\rm CoM}|\Psi_{0 00}^{\rm CoM}} \right|^{2} \,,
\end{equation}
which 
%include one-body matrix elements
%, which can be calculated
%analytically in most cases, 
%and
must fulfill the relation
\begin{equation}
\label{uno}
\sum_{N=0}^{\infty} K_N^{\rm CoM} =  \braket{\Psi_{0 00}^{\rm CoM}|\hat{O}^{\dagger \rm CoM}\hat{O}^{\rm CoM}|\Psi_{0 00}^{\rm CoM}} \, ,
\end{equation}
as the sum of the transition probabilities to all excited states must always be  equal to  1.
 %We note that
%in Eq.~(\ref{transitionprob}) the sum over $J$ represents the sum over all the degenerate states for a given $N$
%\begin{equation}
%D_N = 1 + \frac{2 N - 1 + {(- 1)}^N }{4}\,.
%\end{equation}
% More explicitly, we have that
%$J\in\{\rm mod} (N,2) , {\rm mod} (N,2) + 2 , \ldots , N \}$.

\begin{widetext}
Finally, the total response function (\ref{eq:Separated_response}) can be written as
\begin{equation}\label{eq:respone total with P_N}
R (\omega) = \frac{1}{2 j_0 + 1} \sum_{m_0} \sum_{\varepsilon jm} \left| \braket{\Psi_{\varepsilon jm}^{\rm I}|\hat{O}^{\rm I}|\Psi_{\varepsilon_0 j_0 m_0}^{\rm I}}
\right|^{2}~\sum_N K_{N}^{\rm CoM} \delta \left(\varepsilon - \varepsilon_0 + N \hbar \Omega - \hbar \omega \right) \,.
\end{equation}
At this point, it is useful to introduce the intrinsic response function,
\begin{equation}
R^{\rm I} (\omega) = \frac{1}{2 j_0 + 1} \sum_{m_0} \sum_{\varepsilon jm} \left| \braket{\Psi_{\varepsilon jm}^{\rm I}|\hat{O}^{\rm I}|\Psi_{\varepsilon_0 j_0 m_0}^{\rm I}} \right|^{2}
\delta \left(\varepsilon - \varepsilon_0 - \hbar \omega \right) \,,
\end{equation}
which is the object we are interested in, as it depends only on the internal degrees of freedom of the nucleus and not on the CoM.  By looking at its definition, we can see that  Eq.~(\ref{eq:respone total with P_N}) contains indeed the intrinsic response function, but for a
different energy transfer: $\omega \; \rightarrow \; \omega - N \Omega$, namely
\begin{equation}
R^{\rm I} (\omega - N \Omega) = \frac{1}{2 j_0 + 1} \sum_{m_0} \sum_{\varepsilon jm}
\left| \braket{\Psi_{\varepsilon jm}^{\rm I}|\hat{O}^{\rm I}|\Psi_{\varepsilon_0j_0m_0}^{\rm I}} \right|^{2}
\delta ( \varepsilon-\varepsilon_0 + N \hbar \Omega - \hbar \omega ) \,.
\end{equation}
If the  energy transfer is smaller than the energy of level $N$, then the system cannot be excited to that energy level.
Mathematically, we demand therefore that $\omega-N\Omega\geq0$.
%%\begin{equation}\label{eq:scatterer bigger energy level}
%%\omega-N\Omega\geq0\,.
%%\end{equation}
 Finally, we can connect the total response function with the intrinsic response function and write the following system of equations 
\begin{equation}\label{eq:response final}
R (\omega) = \sum_N K_{N}^{\rm CoM} R^{\rm I} (\omega - N \Omega ) \; , \qquad \omega - N \Omega \geq 0 \, .
\end{equation}
The idea of this formalism is actually to draw $R^{\rm I} (\omega)$ for a known calculation or $R(\omega)$ at any value of $\omega$ in a certain range of interest.
As a consequence of the condition in Eq.~(\ref{eq:response final}),
%%(\ref{eq:scatterer bigger energy level}),
if $\omega < \Omega$ then $N$ must be zero and
 Eq.~(\ref{eq:response final}) is reduced to $R (\omega) = K_0^{\rm CoM} R^{\rm I} (\omega)$.
At higher energy transfer, $\Omega < \omega < 2 \Omega$, the possible energy levels will include
$N = 1$ and so on. Thus, we can write
\begin{equation}\label{omegaintervals}
\begin{split}
0 \le \omega < \Omega \quad \rightarrow \quad N = 0 \;: \quad R (\omega) &= K_{0}^{\rm CoM} R^{\rm I} (\omega) \, , \\
\Omega \le \omega < 2 \Omega \quad \rightarrow \quad N = 1 \;: \quad R (\omega) &= K_0^{\rm CoM} R^{\rm I} (\omega)
+ K_1^{\rm CoM} R^{\rm I} (\omega - \Omega) \, , \\
2 \Omega \le \omega < 3 \Omega \quad \rightarrow \quad N=2 \;: \quad R (\omega) &= K_0^{\rm CoM} R^{\rm I} (\omega)
+ K_1^{\rm CoM} R^{\rm I} (\omega - \Omega) + K_2^{\rm CoM} R^{\rm I} (\omega - 2 \Omega) \, , \\
&\hspace{0.15cm} \vdots \\
N \Omega \le \omega < (N+1) \Omega \; : \quad R (\omega) &= K_0^{\rm CoM} R^{\rm I} (\omega)
+ \sum_{m=1}^N K_m^{\rm CoM} R^{\rm I} (\omega - m \Omega) \, .
\end{split}
\end{equation}
In the first line of Eq.~(\ref{omegaintervals}) we have only one unknown, namely $R^{\rm I} (\omega)$.
Both $R^{\rm I} (\omega)$ and $R^{\rm I} (\omega - \Omega)$ in the second line are unknown, but we can calculate the
latter using the first line. Indeed, this quantity depends on an energy in the region $\omega < \Omega$, that we can calculate
with the first line of Eq.~(\ref{omegaintervals}). After this is done we can solve the second line. We then solve the third line  with the help of the first and second line, and so on.
For a given energy range $N \Omega \le \omega < (N+1) \Omega$, the recursion can be written as
\begin{equation}\label{eq:Responses relation - intrinsic/total}
R^{\rm I} (\omega) = \frac{1}{K_0^{\rm CoM}} \left[ R (\omega) - \sum_{m=1}^N K_{m}^{\rm CoM} R^{\rm I} (\omega
- m \Omega) \right] \, .
\end{equation}
This is the first important recursive equation resulting from our formalism. In essence,
following this procedure for each region of $\omega$ we can find  the  response function $R^{\rm I} (\omega)$ as calculated in the intrinsic frame  from a calculation of $R(\omega)$ in the laboratory frame, given that we have calculated the CoM transition probabilities of Eq.~(\ref{transitionprob}).

%%%%%%%%%%%%%%%%%%%%%%%%%%SECTION %%%%%%%%%%%%%%%%%%%%%%%%%%%%%%%%%%%%%%

\section{Sum Rules}
\label{section:sum_rule}
In this section we apply our formalism to other important quantities in scattering processes, namely the sum
rules. The latter are  defined as  integrals of the response function 
\begin{equation}
S_n = \int_0^{\infty} d \omega \omega^n R (\omega) \,,
\end{equation}
where $n$ is typically an integer number. Sum rules are sometimes easier to calculate than the response function itself, see, e.g., Refs.~\cite{miorelli2016,miorelli2018,Baker2020}.
Let us assume that a generic sum rule of the total system is calculated in the laboratory frame
and use again the known $K_{N}^{\rm CoM}$ to extract only the intrinsic part, which we are interested in.

If we multiply the relation of Eq.~(\ref{eq:Responses relation - intrinsic/total}) by $\omega^{n}$ and take an integral from zero to infinity, we will obtain a relation for the sum rules. Recalling that $K_{N}^{\rm CoM}$
does not depend on $\omega$ and that a response for negative energies is zero, we can write% a generic expression as
\begin{equation}
S_n = \int_0^{\infty} d\omega \omega^n R (\omega)  = K_0^{\rm CoM} \int_0^{\infty} d\omega \omega^n R^{\rm I} (\omega)
+ K_1^{\rm CoM} \int_{\Omega}^{\infty} d\omega \omega^n R^{\rm I} (\omega-\Omega)
+ K_2^{\rm CoM} \int_{2 \Omega}^{\infty} d\omega \omega^n R^{\rm I} (\omega - 2 \Omega)+ \ldots \, .
\end{equation}
Introducing the intrinsic sum rule
\begin{equation}
S^{\rm I}_n = \int_0^{\infty} d\omega \omega^n R^{\rm I} (\omega)
\end{equation}
and redefining the integration variable in each integral
\begin{eqnarray}
&&\omega - \Omega = \omega_1 \quad \rightarrow  \omega = \omega_1 + \Omega \\
\nonumber
&&\omega - 2 \Omega = \omega_2 \quad \rightarrow  \omega = \omega_2 + 2 \Omega \\
\nonumber
&&\vdots   \\
\nonumber
&&\omega - m \Omega = \omega_m \quad \rightarrow  \omega = \omega_m + m \Omega
\end{eqnarray}
we obtain 
\begin{eqnarray}
S_n &=& K_0^{\rm CoM} S_n^{\rm I} + K_1^{\rm CoM} \int_0^{\infty} d \omega_1 R^{\rm I} (\omega_1) (\omega_1 + \Omega)^n
+ K_2^{\rm CoM} \int_0^{\infty} d \omega_2 R^{\rm I} (\omega_2) (\omega_2 + 2 \Omega)^n + \ldots \\
\nonumber
&+ &K_m^{\rm CoM} \int_0^{\infty} d \omega_m R^{\rm I} (\omega_m) (\omega_m + m \Omega)^n + \ldots \, .
\end{eqnarray}
The binomial coefficients $(x+y)^n = \sum_{k=0}^n \binom{n}{k} x^{n-k} y^k$ can be used to rewrite 
\begin{equation}
(m \Omega + \omega_m)^n = \sum_{k=0}^n \binom{n}{k} (m \Omega)^{n-k} \omega_m^k
\end{equation}
and  obtain 
\begin{eqnarray}
S_n &=& K_0^{\rm CoM} S_n^{\rm I} + K_1^{\rm CoM} \int_0^{\infty} d \omega_1 R^{\rm I} (\omega_1) \sum_{k=0}^n \binom{n}{k} (\Omega)^{n-k} \omega_1^k
+  K_2^{\rm CoM} \int_0^{\infty} d \omega_2 R^{\rm I} (\omega_2) \sum_{k=0}^n \binom{n}{k} (2 \Omega)^{n-k} \omega_2^k + \ldots \\
\nonumber
&+&K_m^{\rm CoM} \int_0^{\infty} d \omega_m R^{\rm I} (\omega_m) \sum_{k=0}^n \binom{n}{k} (m \Omega)^{n-k} \omega_m^k + \ldots \\
\nonumber
&=& K_0^{\rm CoM} S_n^{\rm I} + K_1^{\rm CoM} \sum_{k=0}^n \binom{n}{k} (\Omega)^{n-k} \int_0^{\infty} d \omega_1 R^{\rm I} (\omega_1) \omega_1^k
+  K_2^{\rm CoM} \sum_{k=0}^n \binom{n}{k} (2 \Omega)^{n-k} \int_0^{\infty} d \omega_2 R^{\rm I} (\omega_2) \omega_2^k + \ldots \\
\nonumber
&+& K_m^{\rm CoM} \sum_{k=0}^n \binom{n}{k} (m \Omega)^{n-k} \int_0^{\infty} d \omega_m R^{\rm I} (\omega_m) \omega_m^k + \ldots \, .
\end{eqnarray}
Since  the $\omega_m$ are dummy integration variables, we can at this point replace them by $\omega$, so that the integrals become sum rules and we obtain
\begin{eqnarray}
S_n &=& K_0^{\rm CoM} S_n^{\rm I} + K_1^{\rm CoM} \sum_{k=0}^n \binom{n}{k} (\Omega)^{n-k} \int_0^{\infty} d \omega R^{\rm I} (\omega) \omega^k
+ K_2^{\rm CoM} \sum_{k=0}^n \binom{n}{k} (2\Omega)^{n-k} \int_0^{\infty} d \omega R^{\rm I} (\omega) \omega^k + \ldots \\
\nonumber
&+& K_m^{\rm CoM} \sum_{k=0}^n \binom{n}{k} (m \Omega)^{n-k} \int_0^{\infty} d \omega R^{\rm I} (\omega) \omega^k + \ldots \\
\nonumber
&=& K_0^{\rm CoM} S_n^{\rm I} + K_1^{\rm CoM} \sum_{k=0}^n \binom{n}{k} (\Omega)^{n-k} S_k^{\rm I}
+ K_2^{\rm CoM} \sum_{k=0}^n \binom{n}{k} (2 \Omega)^{n-k} S_k^{\rm I} + \ldots + K_m^{\rm CoM} \sum_{k=0}^n \binom{n}{k} (m \Omega)^{n-k} S_k^{\rm I} +\ldots\\
\nonumber
&=& K_0^{\rm CoM} S_n^{\rm I} + \sum_{k=0}^n \binom{n}{k} (\Omega)^{n-k} S_k^{\rm I} \left[K_{1}^{\rm CoM} + K_2^{\rm CoM} 2^{n-k}+ \ldots
+ K_m^{\rm CoM} m^{n-k} + \ldots \right] \, .
\end{eqnarray}
%\end{widetext}
Thus, we can write the following recursive relation for the intrinsic sum rules
\begin{equation}\label{eq:S_n final}
S_n  =  K_0^{\rm CoM} S_n^{\rm I} + \sum_{k=0}^n \binom{n}{k} (\Omega)^{n-k} S_{k}^{\rm I} \sum_{m=1}^{\infty} K_m^{\rm CoM} m^{n-k} \, .
\end{equation}
To obtain intrinsic sum rules from a calculation in the laboratory frame, we clearly have an iterative procedure.  First, we find the intrinsic sum rule of order zero $S_0^{\rm I}$ as
\begin{equation}\label{eq:S_0 intrinsic}
S_0^{\rm I} = \frac{S_0}{\sum_{m=0}^{\infty} K_m^{\rm CoM}} \, .
\end{equation}
Second, to obtain the intrinsic sum rule of order one we use
\begin{eqnarray}
S_1 &=& K_0^{\rm CoM} S_1^{\rm I} + \sum_{k=0}^1 \binom{1}{k} (\Omega)^{1-k} S_k^{\rm I} \sum_{m=1}^{\infty} K_m^{\rm CoM} m^{1-k} \\
\nonumber
&=& S_1^{\rm I} \sum_{m=0}^{\infty} K_m^{\rm CoM} + \Omega S_0^{\rm I} \sum_{m=1}^{\infty} m K_m^{\rm CoM} \, ,
\end{eqnarray}
so that the intrinsic sum rule can be obtained as 
\begin{equation}\label{eq:S_1 intrinsic}
S_1^{\rm I} = \frac{1}{\sum_{m=0}^{\infty} K_m^{\rm CoM}} \left[ S_1 - \Omega S_0^{\rm I} \sum_{m=1}^{\infty} m K_m^{\rm CoM} \right] \, . 
\end{equation}
%\begin{widetext}
Next, for the sum rule of second order we have instead
\begin{eqnarray}
S_2 & =& S_2^{\rm I} \sum_{m=0}^{\infty} K_m^{\rm CoM} + \Omega^2 S_0^{\rm I} \left[ K_1^{\rm CoM} + K_2^{\rm CoM} 2^2 + \ldots
K_m^{\rm CoM} m^2 + \ldots \right] + 2 \Omega S_1^{\rm I} \left[ K_1^{\rm CoM} + K_2^{\rm CoM} 2 + \ldots + K_m^{\rm CoM} m + \ldots \right] \\
\nonumber
&=& S_2^{\rm I} \sum_{m=0}^{\infty} K_m^{\rm CoM} + S_1^{\rm I} 2 \Omega \sum_{m=1}^{\infty} m K_m^{\rm CoM} + S_0^{\rm I}
\Omega^2 \sum_{m=1}^{\infty} m^2 K_m^{\rm CoM} \, ,
\end{eqnarray}
so that the intrinsic sum rule of second order can be obtained as
\begin{equation}\label{eq:S_2 intrinsic}
S_2^{\rm I} = \frac{1}{\sum_{m=0}^{\infty} K_m^{\rm CoM}} \left[S_2 - S_1^{\rm I} 2 \Omega \sum_{m=1}^{\infty} m K_m^{\rm CoM} - S_0^{\rm I} \Omega^2
\sum_{m=1}^{\infty} m^2 K_m^{\rm CoM} \right] \, . 
\end{equation}
Finally, for a generic sum rule of order $n$ we have 
\begin{eqnarray}
S_n &=& K_0^{\rm CoM} S_n^{\rm I} + \sum_{k=0}^n \binom{n}{k} (\Omega)^{n-k} S_k^{\rm I} \sum_{m=1}^{\infty} K_m^{\rm CoM} m^{n-k}
= K_0^{\rm CoM} S_n^{\rm I} + S_n^{\rm I} \sum_{m=1}^{\infty} K_m^{\rm CoM} + \sum_{k=0}^{n-1} \binom{n}{k} (\Omega)^{n-k} S_k^{\rm I}
\sum_{m=1}^{\infty} K_m^{\rm CoM} m^{n-k} \\
\nonumber
&=& S_n^{\rm I} \sum_{m=0}^{\infty} K_m^{\rm CoM} + \sum_{k=0}^{n-1} \binom{n}{k} (\Omega)^{n-k} S_k^{\rm I} \sum_{m=1}^{\infty} K_m^{\rm CoM} m^{n-k} \, ,
\end{eqnarray}
and the intrinsic sum rule of order $n$ can be found as
\begin{equation}\label{eq:S_n intrinsic final}
S_n^{\rm I} = \frac{1}{\sum_{m=0}^{\infty} K_m^{\rm CoM}} \left[ S_n - \sum_{k=0}^{n-1} \binom{n}{k} (\Omega)^{n-k} S_k^{\rm I} \sum_{m=1}^{\infty}
K_m^{\rm CoM} m^{n-k} \right] \, .
\end{equation}
This is the second important recursive equation resulting from our formalism.
\end{widetext}

%%%%%%%%%%%%%%%%%%%%%%%%%%%%%%%%%%%%%% SECTION %%%%%%%%%%%%%%%%%%%%%%%%%%%%%%%%%%%%%%

\section{Electron scattering process}
\label{section:el_scatt}

We now direct our attention to an application of the newly derived formalism to a chosen physical scattering process that serves as an example.
In the inclusive electron scattering off an $A$-body nucleus, one deals with two kind of electromagnetic response functions, namely the longitudinal and the transverse response function. Here we will consider the former, which involves the charge operator
\begin{equation}
\hat{O} = \sum_{k=1}^Z e^{i \mathbf{q} \cdot \mathbf{r}_k} \, ,
\end{equation}
where $\mathbf{r}_k$ is the coordinate of each nucleon in the laboratory frame, $\mathbf{q}$ is the momentum transfer from the  electron to the nucleus and $Z$ is the number of protons in the nucleus.
If we use Eq.~(\ref{rel}),
the charge operator 
\begin{equation}
 \hat{O}= e^{i \mathbf{q} \cdot \mathbf{R}_{\rm CoM}} \sum_{k=1}^Z e^{i \mathbf{q} \cdot \mathbf{r}'_k} \, ,
\end{equation}
factorizes as in Eq.~(\ref{eq:Operator decoupling}) 
with
\begin{equation}\label{coulomboperators}
\hat{O}^{\rm CoM} = e^{i \mathbf{q} \cdot \mathbf{R}_{\rm CoM}} \, , \hspace{1.0cm}
\hat{O}^{\rm I} = \sum_{k=1}^Z e^{i \mathbf{q} \cdot \mathbf{r}'_k} \, .
\end{equation}
 The multipole expansions of the CoM and intrinsic coordinates in plane waves read~\cite{edmonds1996angular}
\begin{eqnarray}
e^{i \mathbf{q} \cdot \mathbf{R}_{\rm CoM}} &= 4 \pi \sum_{JM} i^J j_J (qR_{\rm CoM}) Y_{JM} (\hat{\mathbf{R}}_{\rm CoM})
Y_{JM}^{\star} (\hat{\mathbf{q}}) \nonumber \label{pwcomchargeop} \\
&\equiv \sum_{JM} \hat{C}^{\rm CoM}_{JM} 
\end{eqnarray}
and
\begin{eqnarray}
\sum_{k=1}^Z e^{i \mathbf{q} \cdot \mathbf{r}'_k} &= 4 \pi \sum_{jm} i^j
\sum_{k=1}^Z  j_j (q r'_k) Y_{jm} (\hat{\bf r}'_k) Y_{jm}^{\star} (\hat{\bf q}) \nonumber \\
&\equiv \sum_{jm} \hat{C}^{\rm I}_{jm} \, ,
\end{eqnarray}
where  $\hat{C}^{\rm CoM}_{JM}$ and $\hat{C}^{\rm I}_{jm}$ are the CoM and intrinsic Coulomb multipoles, respectively.
The CoM wave function is a harmonic oscillator state, written in general as 
\begin{equation}
\Psi_{N JM}^{\rm CoM} ({\bf R}_{\rm CoM}) = R_{N_r J} (R_{\rm CoM}) Y_{JM} (\hat{\bf R}_{\rm CoM}),
\end{equation}
with energy $E= \hbar \Omega (N + \frac{3}{2})$ where $N = 2N_r + J$ and $J \le N$.
In the ground state, the oscillator wave function has the quantum numbers $N_r^0=0,J_0=0,M_0=0$  so that the initial state will be 
\begin{equation}
\label{CoM_gs}
\Psi_{0 00}^{\rm CoM} ({\bf R}_{\rm CoM}) = \sqrt{\frac{2}{4\pi b_{\rm CoM}^{3} \Gamma \left(3/2 \right)}}
e^{- \frac{R_{\rm CoM}^{2}}{2b_{\rm CoM}^{2}}} \, ,
\end{equation}
where $b_{\rm CoM}=\sqrt{\frac{\hbar}{ A m_N \Omega}}$ is the harmonic oscillator length. 

Our formalism requires the knowledge of the CoM transition probabilities of Eq.~(\ref{transitionprob}) for the CoM Coulomb operator $\hat{C}^{\rm CoM}_{JM}$. Since it carries angular momentum $J$ with projection $M$ and the quantum numbers of the initial CoM state are all zero, obviously the final state quantum numbers and wave function will be~\cite{-Suhonen-J}) 
\begin{equation}\label{eq:radial final  w.f.}
\begin{split}
\Psi^{\rm CoM}_{N J M}({\bf R}_{\rm CoM}) &= R_{N_r J}(R_{\rm CoM})Y_{JM}(\hat{\bf R}_{\rm CoM}) \\
&= \sqrt{\frac{2N_r !}{b_{\rm CoM}^{3}\Gamma\left(N_r+J+3/2\right)}}\left(\frac{R_{\rm CoM}}{b_{\rm CoM}}\right)^{J} \\
&\times e^{-\frac{{R}_{\rm CoM}^{2}}{2b_{\rm CoM}^{2}}} L_{N_r}^{J+\frac{1}{2}}\left(\frac{R_{\rm CoM}^{2}}{b_{\rm CoM}^{2}}\right) Y_{JM}(\hat{\bf R}_{\rm CoM})\,.
\end{split}
\end{equation}

\begin{figure}[th]
\begin{center}
\includegraphics[scale=0.37]{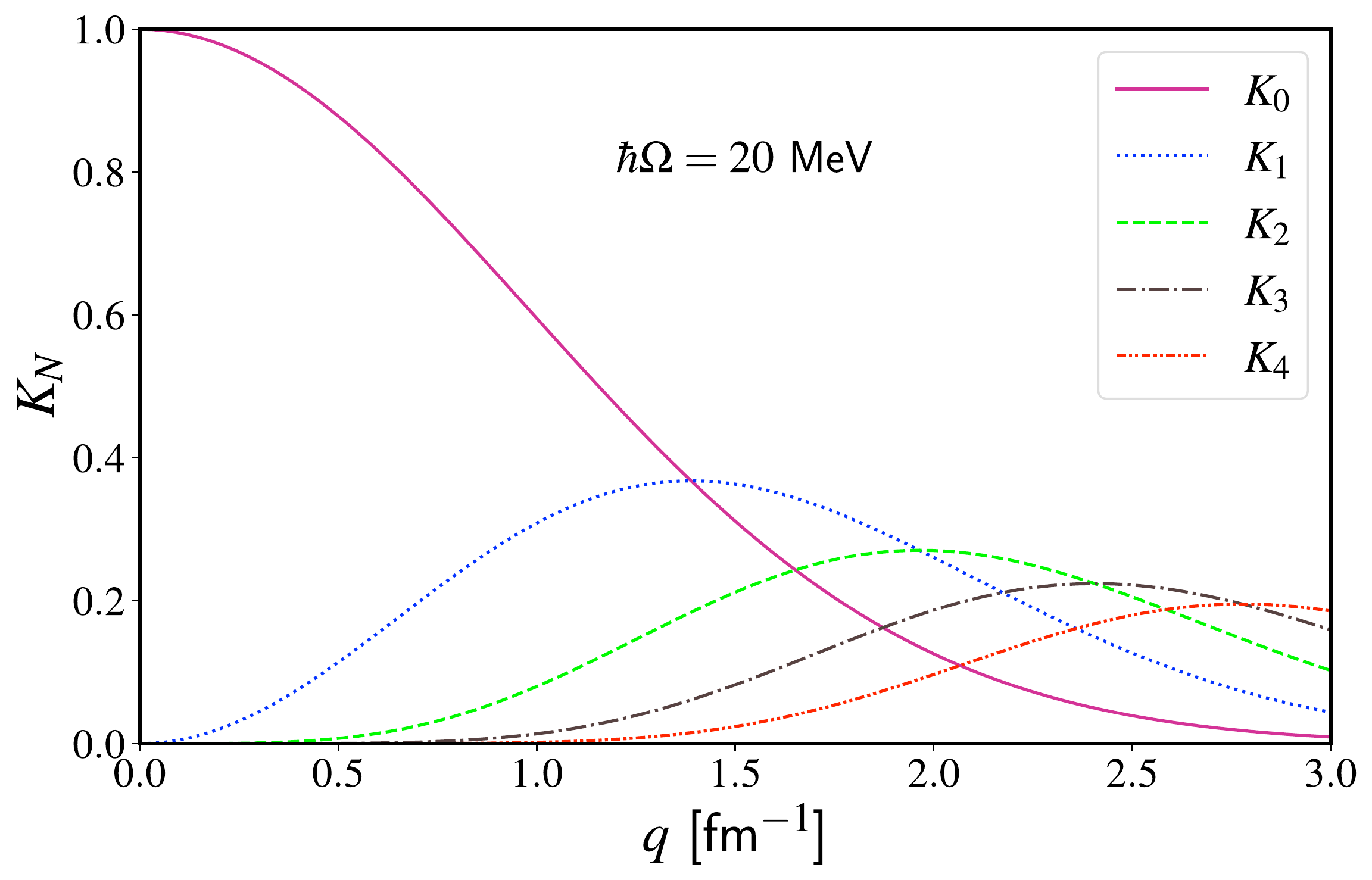} 
\caption{\label{ktansprob} The CoM transition probabilities for the Coulomb operator as a function of the momentum transfer $q$.  The first five values of the quantum number $N$ are shown for $\hbar\Omega=20$ MeV.}
\end{center}
\end{figure}

Starting from Eq.~(\ref{transitionprob}) and using Eq.~(\ref{coulomboperators}) and (\ref{pwcomchargeop})
for $\hat{O}^{\rm CoM}$ we obtain 
\begin{eqnarray}\label{tranprobsec}
K_N^{\rm CoM} &=& \sum_{JM} \left| \braket{\Psi_{N JM}^{\rm CoM}|\hat{O}^{\rm CoM}|\Psi_{000}^{\rm CoM}} \right|^2\\
\nonumber
&= &\sum_{JM} \sum_{J^{\prime} M^{\prime}} \left| \braket{\Psi_{N JM}^{\rm CoM}|\hat{C}^{\rm CoM}_{J^{\prime} M^{\prime}}|\Psi_{000}^{\rm CoM}} \right|^2\\
\nonumber
&=& \sum_{JM} \left| \braket{\Psi_{NJM}^{\rm CoM}|\hat{C}^{\rm CoM}_{JM}|\Psi_{000}^{\rm CoM}} \right|^2 \, ,
\end{eqnarray}
where the sum over $J^{\prime}$ and $M^{\prime}$ is dropped due to selection rules. 
Using the Wigner-Eckart theorem~\cite{edmonds1996angular}, the matrix elements of  $\hat{C}^{\rm CoM}_{JM}$ can be related to that with angular momentum projection $M=0$, so that essentially one only needs $\braket{\Psi_{NJM}^{\rm CoM}|\hat{C}^{\rm CoM}_{J0}|\Psi_{000}^{\rm CoM}}$. The latter turns out to be expressed
in terms of the Gamma's and Kummer's functions ($M$) as~\cite{tianrui,israel}
\begin{eqnarray}
&\braket{\Psi_{NJ0}^{\rm CoM}|\hat{C}^{\rm CoM}_{J0}|\Psi_{000}^{\rm CoM}}=\\
\nonumber
&= i^J \sqrt{\frac{\sqrt{\pi} (2J+1) \Gamma(N_r+J+3/2) \Gamma (N_r + 1)}{2 \Gamma^2 (J + 3/2)}} \left(\frac{b_{\rm CoM} q}{2} \right)^{J} \\
\nonumber
&\times
\sum_{m=0}^{N_r} {(-1)}^{m} \frac{1}{\Gamma (N_r - m + 1) \Gamma (m + 1)} \\
\nonumber
&M \left(J+\frac{3}{2}+m,J+\frac{3}{2},-\frac{b_{\rm CoM}^{2}q^{2}}{4}\right) \, .
\end{eqnarray}
Details for the analytical derivation are found in Appendix~\ref{AppA}.
%, leading to Eq.~(\ref{CoulombME}).
Using Eq.~(\ref{tranprobsec}) and Eq.~(\ref{eq:Analitical solution}) one can write down the first CoM transition probabilities as 
\begin{equation}
K_N^{\rm CoM} = \sum_J \left| \braket{\Psi_{N J0}^{\rm CoM}|\hat{C}^{\rm CoM}_{J0}|\Psi_{000}^{\rm CoM}} \right|^2\,,
\end{equation}
which we plot for the first few $N$ in Fig.~\ref{ktansprob}
 for 
$\hbar \Omega=20$ MeV.  We have numerically verified that Eq.~(\ref{uno}) holds true.
These expressions of the CoM transition probabilities have already been used in Ref.~\cite{Sobczyk2020} to analyze the CoM contamination of coupled-cluster theory calculations of the Coulomb sum rule.

\section{The photoabsorption process}
\label{section:photoabs}

We now turn our attention to another example, namely the photoabsorption process, i.e., the absorption of a real photon by a nucleus. In this case we have a relation between the energy transfer and the momentum transfer, namely $\omega=|\bf{q}|$. In the low-energy regime the leading response function is the so called unretarded dipole 
response function~\cite{gazit2006,bacca2013,bacca2014} and the 
 $z$-component of the dipole operator can be written as
\begin{equation}
\label{dipole}
\hat{D}_z = \sum_{i=1}^Z {z}_k \, ,
\end{equation}
where $z_k$ is the $z$-component of the $k$-th particle
coordinate in the laboratory frame. Clearly, in this case, if we use the relative coordinate $z'_k=z_k -Z_{\rm CoM}$, we are not in a situation in which the operator factorizes as in Eq.~(\ref{eq:Operator decoupling}), but rather we have an operator that is the sum of the CoM and intrinsic part
as 
\begin{equation}\label{operator sum}
\hat{O} = \hat{O}^{\rm I} + \hat{O}^{\rm CoM}\,.
\end{equation}
Hence, we cannot immediately apply our formula in Eqs.~(\ref{eq:Responses relation - intrinsic/total}) and (\ref{eq:S_n intrinsic final}).
It is therefore necessary to extend our formalism.
For this purpose, we write
\begin{eqnarray}\label{operator sum_trick}
\hat{O}^{\rm I} + \hat{O}^{\rm CoM} &=& \frac{d}{d \alpha} e^{\alpha (\hat{O}^{\rm I} + \hat{O}^{\rm CoM})} \Big|_{\alpha =0}\\
\nonumber
&=&\frac{d}{d \alpha} \left( \tilde{O}^I \tilde{O}^{\rm CoM} \right) \Big|_{\alpha =0} 
\end{eqnarray}
and reduce the problem to the ansatz of Eq.~(\ref{eq:Operator decoupling}) $\tilde{O}=\tilde{O}^I \tilde{O}^{\rm CoM}$ with the operators  $\tilde{O}^I$ and $\tilde{O}^{\rm CoM} $ 
that depend on the parameter $\alpha$.

We will now focus on the dipole response function and consider
the nuclear two-body problem ($A=2$) to obtain a proof of principle of our formalism. In case of the deuteron, composed by a proton and a neutron, we only have one charged particle. Its $z$-component is denoted with
 $z_1$, so that the dipole operator of Eq.~(\ref{dipole}) becomes
\begin{equation}
\hat{D}_z = z_1 = \frac{z}{2} + Z_{\rm CoM}  \, .
\end{equation}
We define
\begin{eqnarray}
&\Tilde{O}^{\rm I} \equiv e^{\alpha \frac{z}{2}} \,, \\
\nonumber
&\Tilde{O}^{\rm CoM} \equiv e^{\alpha Z_{\rm CoM}}
\end{eqnarray}
so that these operators become
\begin{eqnarray}\label{dipole_approx}
&\Tilde{O}^{\rm I} & \approx 1+ \alpha \frac{z}{2} = 1 + \alpha z'=
1+\alpha  \hat O^I \,, \\
\nonumber
&\Tilde{O}^{\rm CoM} & \approx 1 + \alpha Z_{\rm CoM} = 1+\alpha \hat O^{\rm CoM}
\end{eqnarray}
for small $\alpha$,
where the relative coordinate in the two-body intrinsic frame is $z' = \frac{z_1-z_2}{2}$ and $Z_{CoM} = \frac{z_1+z_2}{2}$.
In this way, we obtain that
\begin{eqnarray} 
\Tilde{O} & = & \tilde{O}^I \tilde{O}^{\rm CoM} \approx \left(1+\alpha \frac{z_1-z_2}{2} \right)\left(1+\alpha \frac{z_1+z_2}{2} \right)  
\\
\nonumber
&=&1 + \frac{\alpha}{2} z_1 + \frac{\alpha^2}{4} \left(z_1^2-z_2^2\right).
\end{eqnarray}

Considering the operators  $\Tilde {O}^I$ in the intrinsic frame and  $\Tilde{O}$ in the laboratory frame, we define the respective sum rules of order $n$ as  $\Tilde{S}_n^I$ and $\Tilde{S}_n$.
From Eq.~(\ref{eq:S_n intrinsic final}),
we obtain that the relations between the  sum rule of the first three orders ($n=0,1,2$) are
\begin{eqnarray}
    \Tilde{S}_0^I &=& \frac{\Tilde{S}_0}{\sum_{N=0}^\infty \Tilde{K}_N^{\rm CoM}}\,, \\
    \nonumber
     \Tilde{S}_1^I &=& \Tilde{S}_1 - \Omega \Tilde{S}^I_0 \sum_{N=0}^\infty \Tilde{K}_N^{\rm CoM}\,,\\
     \nonumber
     \Tilde{S}_2^I &=& \Tilde{S}_2 -2 \Omega \Tilde{S}^I_1 \sum_{N=0}^\infty N \Tilde{K}_N^{\rm CoM} - \Omega^2 \Tilde{S}^I_0 \sum_{N=0}^\infty N^2 \Tilde{K}_N^{\rm CoM}\,.
\end{eqnarray}

We now calculate the CoM transition probabilities $\Tilde{K}_N^{\rm CoM}$.
Details are worked out in Appendix~\ref{AppB}, leading to
\begin{eqnarray}\label{K_CoM_n_D}
\Tilde{K}_0^{\rm CoM} &= &\left| \braket{\Psi_{000}^{\rm CoM}|\Tilde{O}^{\rm CoM}|\Psi_{000}^{\rm CoM}} \right|^2 = 1 \, , \\
\nonumber
\Tilde{K}_1^{\rm CoM} &=& \left| \braket{\Psi_{010}^{\rm CoM}|\Tilde{O}^{\rm CoM}|\Psi_{000}^{\rm CoM}} \right|^2 = \frac{\alpha^2 b_{\rm CoM}^2}{2} \, , \\
\nonumber
\Tilde{K}_N^{\rm CoM} &=& 0 \, , ~{\rm for}~ \quad N \geq 2 \, ,
\end{eqnarray}
where here again $b_{\rm CoM}$ is the CoM harmonic oscillator length 
(this time with mass equal to $2m_N$). We observe then that
\begin{align}\label{K_CoM Tilde}
    \sum_{N=0}^\infty \Tilde{K}_N^{\rm CoM} = 1 + \frac{\alpha^2b_{\rm CoM}^2}{2}\,.
\end{align}
With the help of Eq.~(\ref{K_CoM Tilde}), the relation between the sum rules of order $n=0$ becomes
\begin{align}\label{expected_ratio_0}
    &\frac{\Tilde{S}_0}{\Tilde{S}_0^I} = 1 + \frac{\alpha^2b_{\rm CoM}^2}{2}\,, 
\end{align}
which is the first equation we want to verify.

In this way, using  again Eq.~(\ref{K_CoM_n_D}), we get 
the following relations for the
intrinsic sum rules of order $n = 1,2$ 
\begin{align}\label{expected_relation_1}
\Tilde{S}_1^I = \frac{1}{1+\frac{\alpha^2b_{\rm CoM}^2}{2}} \left[\Tilde{S}_1 - \Omega \Tilde{S}_0^I\frac{\alpha^2b_{\rm CoM}^2}{2} \right]
\end{align}
and 
\begin{align}\label{expected_relation_2}
\Tilde{S}_2^I =  \frac{1}{1+\frac{\alpha^2b_{\rm CoM}^2}{2}} \left[\Tilde{S}_2 - 2\Omega \Tilde{S}_1^I\frac{\alpha^2b_{\rm CoM}^2}{2}-\Omega^2\Tilde{S}_0^I\frac{\alpha^2b_{\rm CoM}^2}{2}\right].
\end{align}

The intrinsic operator $\tilde{O}^I$ is simply related to the standard  (translational invariant) unretarded dipole operator $\hat{O}^I= z'$,
see \eqref{dipole_approx}. Denoting with $S^I_n$ the corresponding sum
 rules of order $n$, it is not difficult to establish the following
 relations
\begin{eqnarray}
\nonumber
  \Tilde{S}_0^I &=  1 + \alpha ^2 S_0^I~{\rm for}~ n=0\\
  \label{eq_last_fig}
  \Tilde{S}^I_n &= \alpha^2 S_n^I ~{\rm for~higher}~ n\,.
\end{eqnarray}
These are the other two equations we want to verify.

\subsection{Numerical implementation}
At this point we can verify our formalism  with a numerical implementation by checking Eqs.~(\ref{expected_ratio_0}) and (\ref{eq_last_fig}), where for the l.h.s.~of the latter we will exploit (\ref{expected_relation_1}) and (\ref{expected_relation_2}).
To this purpose we use a modern nucleon-nucleon interaction stemming from chiral effective field theory at next-to-next-to-next-to-leading order~\cite{entem2003}, called N$^3$LO-EM, and solve the two-body problem using a harmonic oscillator basis. We use either  the intrinsic frame, where the task reduces to the solution of a one-body problem in the relative coordinate, or in the laboratory frame where we have a two-body problem in the two single-particle coordinates. 

\begin{figure}[th]
 \begin{center}
\includegraphics[scale=0.29]{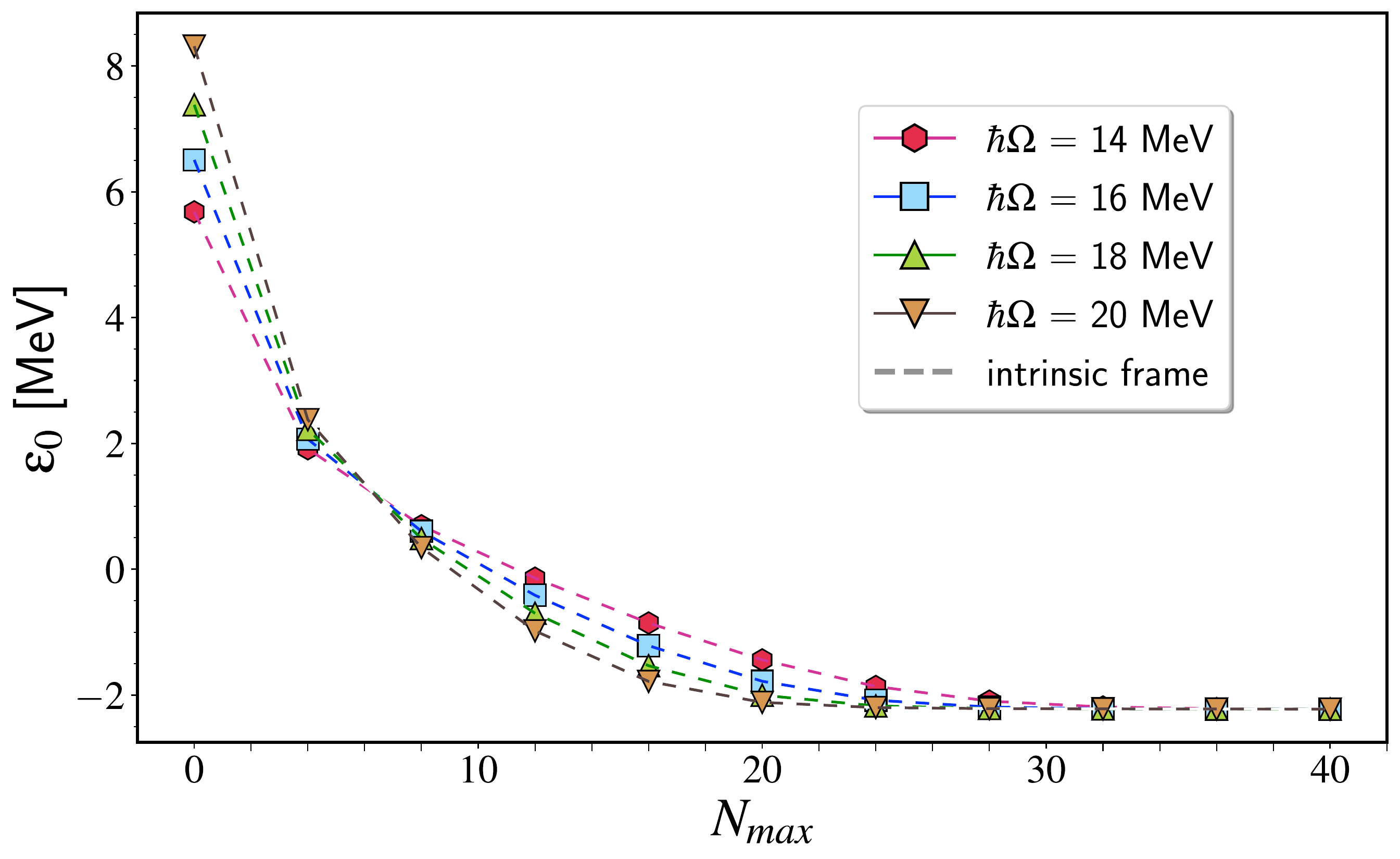} 
 \caption{\label{Plot: binding energy}  Binding energy of the deuteron $\varepsilon_0$ computed in the intrinsic frame and in the laboratory frame as $\mathcal{E}_0 - E_0$ with $E_0 = \frac{3}{2}\hbar \Omega$ as a function of $N_{max}$.}
\end{center}
\end{figure}

Before dealing with our new formalism, we first check our numerical implementation, computing the deuteron ground-state energy  $\varepsilon_0$. In Fig. \ref{Plot: binding energy}, we show
 the  energy calculated in the intrinsic frame and laboratory frame for $N_{max}$ up to 40 and different $\hbar\Omega$. We get a perfect match between both calculations for every model space. As expected we see that, as  $N_{max}$ grows, the energies approach the converged values from above due to the variational principle. At $N_{max}=40$ for $\hbar\Omega=16$ MeV we obtain 
 $\varepsilon_0=-2.2236$ MeV, which is rather close to the 
 experimental value $\varepsilon^{\rm exp}_0 = - 2.224573(2)$ MeV \cite{AUDI20033}. However, the goal of this paper is not necessarily to compare with experiment, but rather to compare calculations in the intrinsic~\cite{dennis} and laboratory frames~\cite{kerem}.
 
\begin{figure}[th]
 \begin{center}
\includegraphics[scale=0.32]{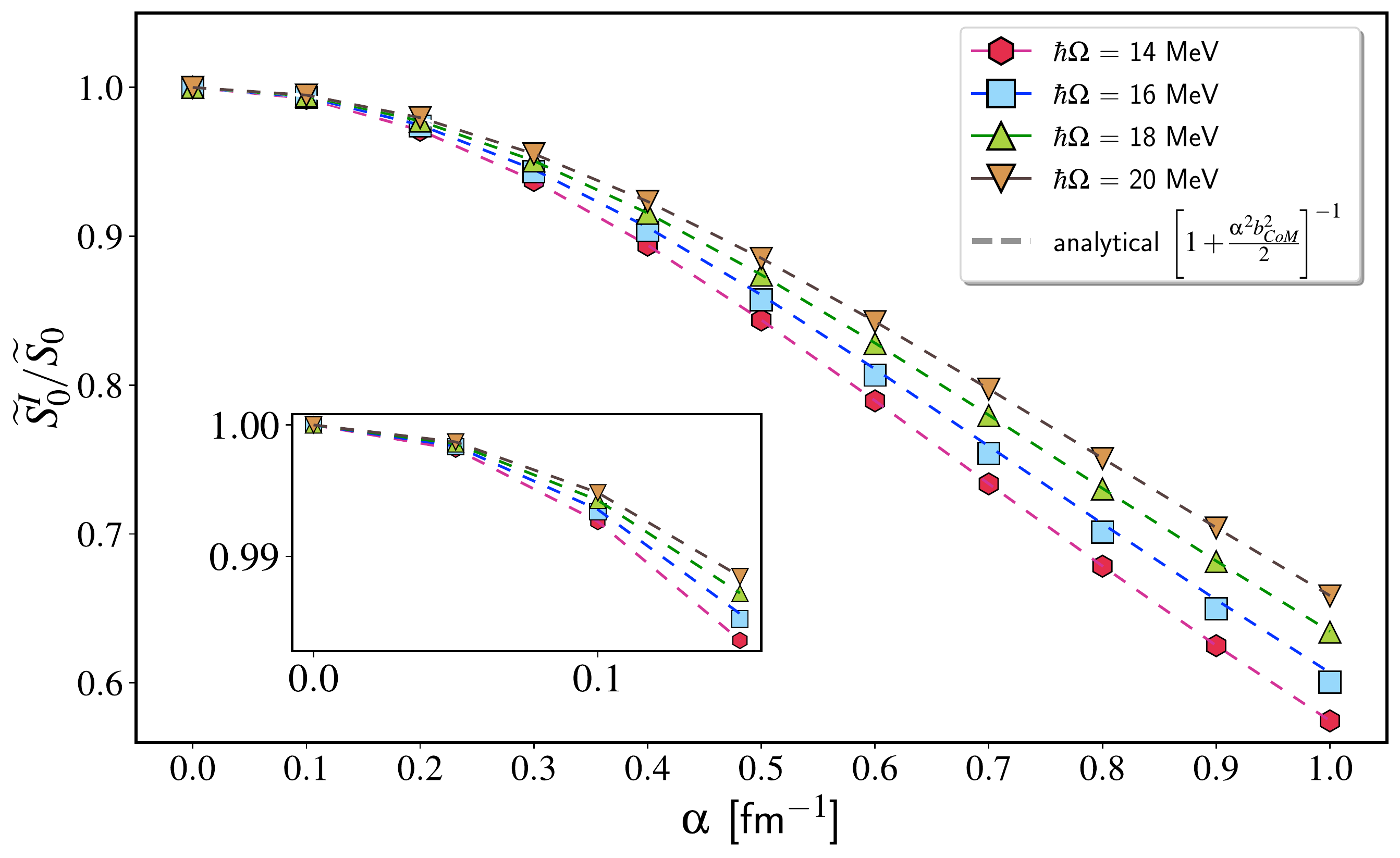} 
 \caption{\label{fig_rel0} Ratio of $\Tilde{S}_0$ and $\Tilde{S}_0^I$ as a function of   $\alpha$ for $N_{max} = 30$ and various $\hbar\Omega$. The expected analytical behavior is shown by the dashed line.}
\end{center}
\end{figure}

Next we implement the dipole operators~\cite{kerem} so we are ready to numerically verify our formalism. 
In Fig.~\ref{fig_rel0}, we plot the l.h.s.~of Eq.~(\ref{expected_ratio_0}), calculated numerically, against the expected analytical behavior.
For every $\alpha$ we obtain agreement  within $10^{-3}-10^{-4}$. To get an ever better agreement one needs to increase the model space size $N_{max}$ so that the sum rules are fully converged. However, in the laboratory frame the size of the Hamiltonian matrix is very large (few GB  for $N_{max}=40$ compared to few MB in the intrinsic frame for $N_{max}=300$), hence  we stop at  $N_{max}=40$ since
the test is sufficient to prove that Eq.~(\ref{expected_ratio_0}) is numerically correct.

\begin{figure}[th]
 \begin{center}
 \includegraphics[scale=0.32]{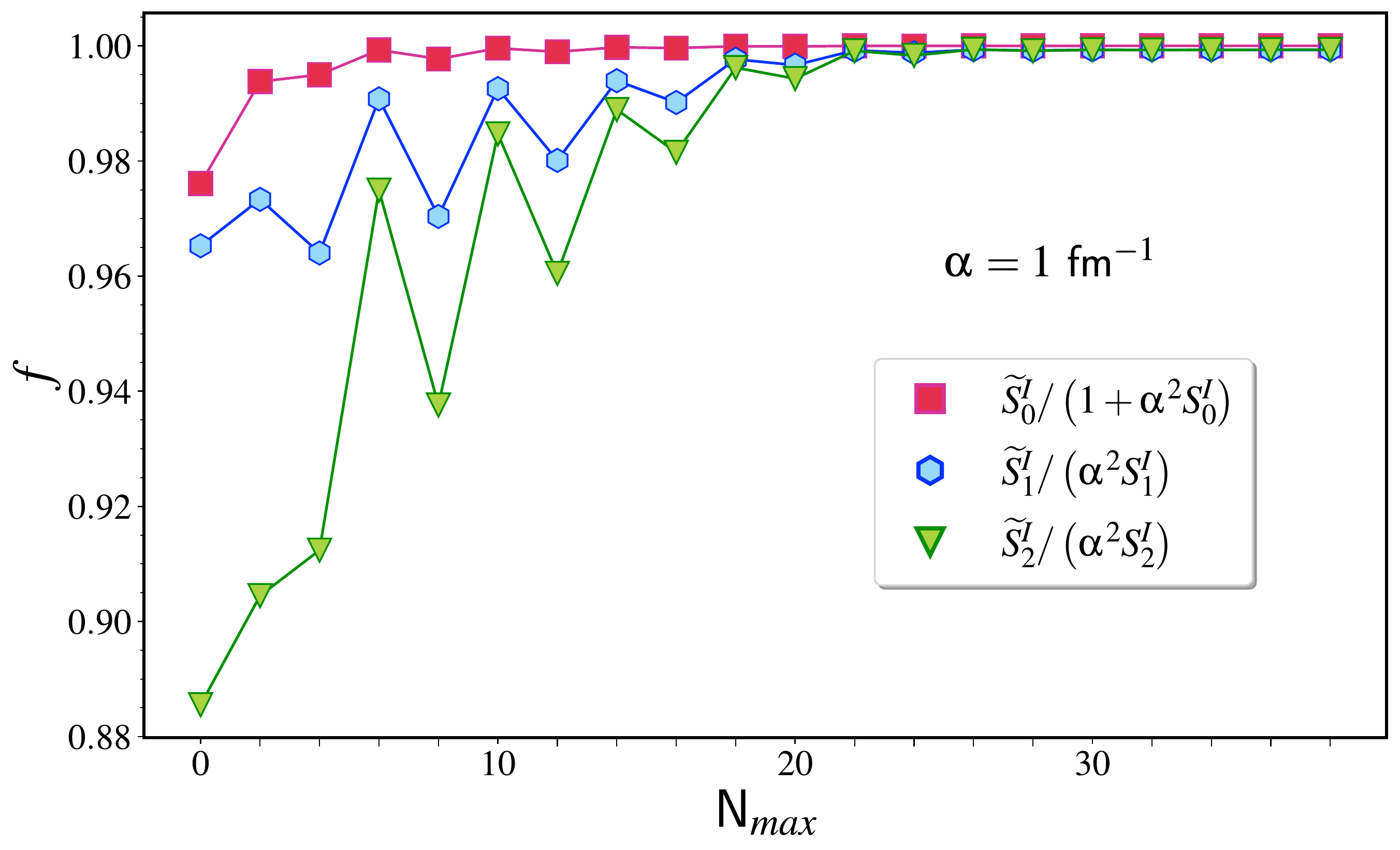} 
 \caption{\label{fig_rel1_2} Ratio $f$ of the l.h.s. and r.h.s. of Eq.~(\ref{eq_last_fig}) as a function various $N_{max}$  for  $\alpha = 1$ fm$^-1$ and  $\hbar\Omega=20$ MeV.} 
\end{center}
\end{figure}

Next, for the sum rules of order $n =0,1,2$, we want to verify Eq.~(\ref{eq_last_fig}). To this purpose,
we divide the l.h.s. by the r.h.s. of this equation  and plot the ratio $f=\frac{\rm l.h.s.}{\rm r.h.s.}$ as a function of  $N_{max}$ in Fig.~\ref{fig_rel1_2}. We clearly see that the ratio converges to 1
when the model space gets large enough, proving that Eq.~(\ref{expected_relation_1}) and (\ref{expected_relation_2}) are numerically verified.
%From the intrinsic frame calculations, we see that sum rules converge after  $N_{max} \approx 50-100$~\cite{dennis}. However, in the laboratory frame the Hamiltonian matrix is too large so we stop at $N_{max} = 40$. This model space is definitely sufficient to prove  the relations in Eq.~(\ref{expected_relation_1}) and (\ref{expected_relation_2}), which is our goal.
This constitutes a proof of principle that the relations we derived are correct. Therefore they can be used for other operators than the dipole in order to extract intrinsic properties from laboratory frame calculations.

\section{Conclusions}
\label{section:conclusions}
In this paper we derive a new formalism to extract intrinsic response functions and sum rules from calculations performed in a general frame. The derivation is based on the use of a harmonic oscillator to confine the CoM motion of the system, a common practice when using a harmonic oscillator single-particle basis, which is at the core of most modern nuclear many-body calculations. The two most relevant recursive relations that we derived are Eq.~(\ref{eq:Responses relation - intrinsic/total}) and Eq.~(\ref{eq:S_n intrinsic final}) for the response function and  for the sum rules, respectively.
After the formal derivation, we discuss two specific examples where this new formalism can be applied. First, we deal with the  case of the electron scattering where we explicitly derive the analytical CoM transition functions.  Second, we analyze the photoabsorption process for which we also provide a numerical implementation in a two-body problem. To be specific,  this allowed us to verify Eq.~(\ref{eq:S_n intrinsic final}) as a proof of principle.
A practical verification of Eq.~(\ref{eq:Responses relation - intrinsic/total}) is left to future work. While one may suspect that  Eq.~(\ref{eq:Responses relation - intrinsic/total}) could get numerically unstable for very large energies $\hbar \Omega$, we expect this relation to be practically usable in the low-energy excitation regime 
of about 20 MeV, which is the range of the giant dipole resonance.
 
 Finally, we would like to point out that this formalism may be of great benefit for calculations of electroweak properties  for nuclei beyond $A=2$, in particular in the $s-$ and $p-$shell nuclei, which owing to their light mass, are mostly affected by the center of mass effects. Applications to other nuclei  will be explored in the future.

\begin{acknowledgments}
S.B. would like to thank Matteo Vorabbi for useful discussions. This work was supported by the Deutsche Forschungsgemeinschaft (DFG) through the  Cluster of Excellence ``Precision Physics, Fundamental
Interactions, and Structure of Matter" (PRISMA$^+$ EXC 2118/1) funded by the
DFG within the German Excellence Strategy (Project ID 39083149), and by the Israel Science Foundation, grant number 1086/21.
\end{acknowledgments}

\appendix
\begin{widetext}
\section{CoM transition matrix elements for the Coulomb operator}
\label{AppA}

Below we show the derivation of the CoM transition matrix element in case of the Coulomb operator, see Eq.~(\ref{coulomboperators}) and (\ref{pwcomchargeop}).
Using the same initial and final wave functions of Eqs.~(\ref{CoM_gs}) and (\ref{eq:radial final  w.f.}) one can analytically calculate the CoM matrix element
as~\cite{israel}
\begin{eqnarray}\label{eq:Integral_full}
\braket{\Psi_{N JM}^{\rm CoM}|\hat{C}^{\rm CoM}_{JM}|\Psi_{000}^{\rm CoM}} &= &\iiint dR_{\rm CoM}\sin\theta d\theta d\varphi  R_{\rm CoM}^{2} \sqrt{\frac{2N_r !}{b_{\rm CoM}^{3}
\Gamma\left(N_r +J+3/2\right)}}\left(\frac{R_{\rm CoM}}{b_{\rm CoM}}\right)^{J}e^{-\frac{R_{\rm CoM}^{2}}{2b_{\rm CoM}^{2}}} \\
\nonumber
&\times& L_{N_r}^{J+\frac{1}{2}} \left(\frac{R_{\rm CoM}^{2}}{b_{\rm CoM}^{2}}\right) Y_{JM}^{\star} (\hat{\mathbf{R}}_{\rm CoM}) \,
4\pi i^{J}j_{J}\left(qR_{\rm CoM}\right)Y_{JM}\left(\hat{\mathbf{R}}_{\rm CoM}\right)Y_{JM}^{\star}\left(\hat{\mathbf{q}}\right)
 \sqrt{\frac{2}{4\pi b_{\rm CoM}^{3}\Gamma\left(3/2\right)}}e^{-\frac{R_{\rm CoM}^{2}}{2b_{\rm CoM}^{2}}} \\
 \nonumber
&=& \frac{4}{b_{\rm CoM}^{3}}\sqrt{\frac{\pi N_r !}{\Gamma(N_r+J+3/2)\Gamma(3/2)}}i^{J}Y_{JM}^{\star}(\hat{\mathbf{q}})
\iint \sin\theta d\theta d\varphi Y_{JM}^{\star}(\hat{\mathbf{R}}_{\rm CoM})Y_{JM}(\hat{\mathbf{R}}_{\rm CoM}) \\
\nonumber
&\times& \int dR_{\rm CoM} R_{\rm CoM}^{2}\left(\frac{R_{\rm CoM}}{b_{\rm CoM}}\right)^{J}e^{-\frac{R_{\rm CoM}^{2}}{b_{\rm CoM}^{2}}}L_{N_r}^{J+\frac{1}{2}}\left(\frac{R_{\rm CoM}^{2}}{b_{\rm CoM}^{2}}\right)j_{J}\left(qR_{\rm CoM}\right) \,.
\end{eqnarray}
The angular part of the integral is exactly the orthogonality relation of spherical harmonics, thus equal to 1. The spherical Bessel function can be written in terms of the
Bessel function~\cite{Abramowitz}
\begin{equation}
j_J (qR_{\rm CoM}) = \sqrt{\frac{\pi}{2q}} R_{\rm CoM}^{-\frac{1}{2}} J_{J+\frac{1}{2}} (qR_{\rm CoM}) \, .
\end{equation}
The Laguerre polynomials are finite degree polynomials expressed by~\cite{Abramowitz} 
\begin{equation}
L_{N}^{\alpha} (x)  = \sum_{m=0}^{N} (-)^{m} \binom{N_r + \alpha}{N - m} \frac{1}{m!} x^{m} \, ,
\end{equation}
so that 
\begin{equation}
L_{N_r}^{J + \frac{1}{2}} \left( \frac{R_{\rm CoM}^{2}}{b_{\rm CoM}^{2}} \right) = \sum_{m=0}^{N_r} (-)^{m} \binom{N_r + J + \frac{1}{2}}{N_r - m}
\frac{1}{m!}\left(\frac{R_{\rm CoM}^{2}}{b_{\rm CoM}^{2}}\right)^{m} \,.
\end{equation}
Substituting the latter into our expression for the matrix element we obtain
\begin{eqnarray}\label{me}
\braket{\Psi_{NJM}^{\rm CoM}|C^{\rm CoM}_{JM}|\Psi_{000}^{\rm CoM}} &=& \frac{4}{b_{\rm CoM}^{3}} \sqrt{\frac{\pi N_r !}{\Gamma(N_r+J+3/2) \Gamma(3/2)}} i^J
Y_{JM}^{\star} (\hat{\mathbf{q}}) \!\! \int \!\!\!\! dR_{\rm CoM} R_{\rm CoM}^2 \left(\frac{R_{\rm CoM}}{b_{\rm CoM}}\right)^{J} \!\! e^{-\frac{R_{\rm CoM}^{2}}{b_{\rm CoM}^{2}}} \\
\nonumber
&\times& \left[\sum_{m=0}^{N_r} (-)^m \binom{N_r +J+\frac{1}{2}}{N_r -m} \frac{1}{m!} \left(\frac{R_{\rm CoM}^{2}}{b_{\rm CoM}^{2}}\right)^m \right]
\left[\sqrt{\frac{\pi}{2q}}R_{\rm CoM}^{-\frac{1}{2}}J_{J+\frac{1}{2}}(qR_{\rm CoM})\right] \\
\nonumber
&=& \frac{4}{b_{\rm CoM}^{3}} \sqrt{\frac{\pi N_r !}{\Gamma(N_r+J+3/2)\Gamma(3/2)}} i^J Y_{JM}^{\star} (\hat{\mathbf{q}}) \sum_{m=0}^{N_r} (-)^m
\binom{N_r +J+\frac{1}{2}}{N_r -m} \frac{1}{m!} b_{\rm CoM}^{-2m-J} \sqrt{\frac{\pi}{2q}} \\
\nonumber
&\times & \int dR_{\rm CoM} R_{\rm CoM}^2 R_{\rm CoM}^{J} R_{\rm CoM}^{-\frac{1}{2}} R_{\rm CoM}^{2m} e^{-\frac{R_{\rm CoM}^{2}}{b_{\rm CoM}^{2}}} J_{J+\frac{1}{2}} (qR_{\rm CoM}) \, .
\end{eqnarray}
To calculate the integral 
\begin{equation}
\label{eq:Radial Integral without constants}
\int dR_{\rm CoM} R_{\rm CoM}^{J+\frac{3}{2}+2m} e^{-\frac{R_{\rm CoM}^{2}}{b_{\rm CoM}^{2}}} J_{J+\frac{1}{2}} (qR_{\rm CoM})
\end{equation}
we can use equation (11.4.28) in Ref.~\cite{Abramowitz}
\begin{equation}
\int_0^{\infty} x^{\mu-1} e^{- a^2 x^2} J_{\nu} (cx) dx = \frac{\Gamma \left(\frac{1}{2} \nu + \frac{1}{2} \mu \right) \left(\frac{1}{2} \frac{c}{a} \right)^{\nu}}{2a^{\mu} \Gamma(\nu+1)}
M \left(\frac{1}{2} \nu + \frac{1}{2} \mu , \nu+1 , - \frac{c^{2}}{4a^{2}} \right) \, .
\end{equation}
The Kummer's function $M$ is
%\footnote{Another notation for this function is $_{1}F_{1}(a,b,x)$%
%} $M\left(a,b,x\right)$ expressed by an infinite series (equation (13.1.2)
%there) 
\begin{equation}
M \left(a,b,x\right) = 1 + \frac{ax}{b} + \frac{a(a+1) x^{2}}{b(b+1)2!} + \frac{(a)_{3}x^{3}}{(b)_{3}3!} + \ldots + \frac{(a)_{n}x^{n}}{(b)_{n}n!} + \ldots \, ,
\end{equation}
where $(a)_{n} = a (a+1) (a+2) \ldots (a+n-1)$ and similarly $(b)_n$.
We define 
\begin{equation}
\mu \equiv J + \frac{5}{2} + 2m \, , \qquad a \equiv \frac{1}{b} \, , \qquad \nu \equiv J + \frac{1}{2} \, , \qquad c \equiv q \, , \qquad x \equiv R_{\rm CoM} \, ,
\end{equation}
and we can write the result of Eq.~(\ref{eq:Radial Integral without constants}) as
\begin{equation}
\int d R_{\rm CoM} R_{\rm CoM}^{J+\frac{3}{2}+2m} e^{-\frac{R_{\rm CoM}^{2}}{b^{2}}} J_{J+\frac{1}{2}} (qR_{\rm CoM})
= \frac{\Gamma(J + \frac{3}{2} + m) \left(\frac{1}{2} q b_{\rm CoM} \right)^{J+\frac{1}{2}}}{2b_{\rm CoM}^{-J-\frac{5}{2}-2m} \Gamma(J+\frac{3}{2})}
\times M \left(J + \frac{3}{2} + m , J + \frac{3}{2} , - \frac{b_{\rm CoM}^{2}q^{2}}{4} \right) \, .
\end{equation}
Finally, we collect all the terms and rewrite Eq.~(\ref{me}) obtaining the following expression
\begin{eqnarray}
\braket{\Psi_{N JM}^{\rm CoM}|C^{\rm CoM}_{JM}|\Psi_{000}^{\rm CoM}} &=& \frac{4}{b_{\rm CoM}^{3}} \sqrt{\frac{\pi N_r !}{\Gamma(N_r +J+3/2) \Gamma(3/2)}} i^J
Y_{JM}^{\star} (\hat{\mathbf{q}}) \sum_{m=0}^{N_r} (-)^m \binom{N_r+J+\frac{1}{2}}{N_r-m} \frac{1}{m!} b_{\rm CoM}^{-2m-J} \\
\nonumber
&\times & \sqrt{\frac{\pi}{2q}} \frac{\Gamma(J+\frac{3}{2}+m) (\frac{1}{2}qb_{\rm CoM})^J (\frac{1}{2} qb)^{\frac{1}{2}}}{2b_{\rm CoM}^{-J-\frac{5}{2}-2m} \Gamma(J+\frac{3}{2})}
M \left(J+\frac{3}{2}+m,J+\frac{3}{2},-\frac{b_{\rm CoM}^{2}q^{2}}{4}\right) \, .
\end{eqnarray}
Thus, the CoM matrix element of the Coulomb multipole becomes
\begin{displaymath}
\begin{split}
\braket{\Psi_{NJM}^{\rm CoM}|C^{\rm CoM}_{JM}|\Psi_{000}^{\rm CoM}} &= i^J \pi \sqrt{\frac{N_r!}{\Gamma(N_r+J+3/2)\Gamma(3/2)}}
\left(\frac{1}{2}qb_{\rm CoM}\right)^J Y_{JM}^{\star} (\hat{\mathbf{q}}) \\
&\times \sum_{m=0}^{N_r} (-)^m \binom{N_r+J+\frac{1}{2}}{N_r-m} \frac{\Gamma(J+\frac{3}{2}+m)}{\Gamma(J+\frac{3}{2})m!}
M \left(J+\frac{3}{2}+m,J+\frac{3}{2},-\frac{b_{\rm CoM}^{2}q^{2}}{4}\right) \, .
\end{split}
\end{displaymath}
%%%%
We are free to choose the coordinate system parallel to the scattering direction $\hat{\mathbf{q}}$, therefore the arguments of the spherical harmonic will
be $\theta=\varphi=0$. 
%This choice vanishes all terms with $M\ne0$ since all of them contains $\sin\theta$. In addition the complex part $e^{iM\varphi}$ equal zero.
Knowing that
\begin{displaymath}
Y_{JM}^{\star} (0,0) = \sqrt{\frac{2 J + 1}{4 \pi}} \delta_{M 0} \, ,
\end{displaymath}
we can proceed only with $M=0$, obtaining
\begin{equation}\label{eq:Analitical solution}
\begin{split}
\braket{\Psi_{NJ0}^{\rm CoM}|C^{\rm CoM}_{J0}|\Psi_{000}^{\rm CoM}} &= i^J \sqrt{\frac{N_r!}{\Gamma(N_r+J+3/2)\Gamma(3/2)}}
\left(\frac{1}{2}qb_{\rm CoM}\right)^{J}\sqrt{\frac{\pi (2J+1)}{4}} \\
&\times \sum_{m=0}^{N_r} (-)^{m} \binom{N_r+J+\frac{1}{2}}{N_r-m} \frac{\Gamma(J+\frac{3}{2}+m)}{\Gamma(J+\frac{3}{2})m!}
M \left(J+\frac{3}{2}+m,J+\frac{3}{2},-\frac{b_{\rm CoM}^{2}q^{2}}{4}\right) \, .
\end{split}
\end{equation}
Note that the phase $i^J$ is not important since we are looking at the squared modulus.
Finally, using the properties of the gamma function we obtain
\begin{equation}
\begin{split}
\label{CoulombME}
\braket{\Psi_{NJ0}^{\rm CoM}|C^{\rm CoM}_{J0}|\Psi_{000}^{\rm CoM}}
&= i^J \sqrt{\frac{\sqrt{\pi} (2J+1) \Gamma(N_r+J+3/2) \Gamma (N_r + 1)}{2 \Gamma^2 (J + 3/2)}} \left(\frac{b_{\rm CoM} q}{2} \right)^{J} \\
&\times \sum_{m=0}^{N_r} {(-1)}^{m} \frac{1}{\Gamma (N_r - m + 1) \Gamma (m + 1)} M \left(J+\frac{3}{2}+m,J+\frac{3}{2},-\frac{b_{\rm CoM}^{2}q^{2}}{4}\right) \, .
\end{split}
\end{equation}

\section{CoM transition matrix elements for the dipole operator}
\label{AppB}

Below we show the derivation of the CoM transition matrix element in case of the $\alpha$-dependent operator of Eq.~(\ref{dipole_approx})
\begin{equation}
K_N^{\rm CoM} = \sum_{JM} \left| \braket{\Psi_{N JM}^{\rm CoM}|\Tilde{O}^{\rm CoM}|\Psi_{000}^{\rm CoM}} \right|^2 \, ,
\end{equation}
with
\begin{equation}
\Tilde{O}^{\rm CoM} \simeq 1 + \alpha \sqrt{\frac{4 \pi}{3}} R_{\rm CoM} Y_{10} (\hat{\mathbf{R}}_{\rm CoM}) \, .
\end{equation}
The CoM matrix element becomes
\begin{equation}
\begin{split}
\braket{\Psi_{N JM}^{\rm CoM}|\Tilde{O}^{\rm CoM}|\Psi_{000}^{\rm CoM}} &= \int_0^{\infty} d R_{\rm CoM} R_{\rm CoM}^2 \int d \hat{\mathbf{R}}_{\rm CoM}
\sqrt{\frac{2 N_r!}{b_{\rm CoM}^3 \Gamma (N_r + J + 3/2)}} {\left( \frac{R_{\rm CoM}}{b_{\rm CoM}} \right)}^J e^{-\frac{R_{\rm CoM}^2}{2 b_{\rm CoM}^2}}
L_{N_r}^{J+\frac{1}{2}} \left( \frac{R_{\rm CoM}^2}{b_{\rm CoM}^2} \right) \\
&\times Y_{JM}^{\star} (\hat{\mathbf{R}}_{\rm CoM}) \left[ 1 + \alpha \sqrt{\frac{4 \pi}{3}} R_{\rm CoM} Y_{10} (\hat{\mathbf{R}}_{\rm CoM}) \right]
\sqrt{\frac{2}{b_{\rm CoM}^3 \Gamma (3/2)}} e^{-\frac{R_{\rm CoM}^2}{2 b_{\rm CoM}^2}} Y_{00} (\hat{\mathbf{R}}_{\rm CoM}) \\
&= \sqrt{\frac{4 N_r!}{b_{\rm CoM}^6 \Gamma (3/2) \Gamma (N_r + J + 3/2)}}  \int_0^{\infty} d R_{\rm CoM} R_{\rm CoM}^2 {\left( \frac{R_{\rm CoM}}{b_{\rm CoM}} \right)}^J
e^{-\frac{R_{\rm CoM}^2}{b_{\rm CoM}^2}} L_{N_r}^{J+\frac{1}{2}} \left( \frac{R_{\rm CoM}^2}{b_{\rm CoM}^2} \right) \\
&\times \left(\delta_{J0} \delta_{M0} + \frac{\alpha}{\sqrt{3}} R_{\rm CoM} \delta_{J1} \delta_{M0}\right) \\
&= \sqrt{\frac{4 N_r!}{b_{\rm CoM}^6 \Gamma (3/2) \Gamma (N_r + J + 3/2)}} \sum_{m=0}^{N_r} (-1)^{m} \binom{N_r + J + \frac{1}{2}}{N_r - m} \frac{1}{m!} b_{\rm CoM}^{-J-2m} \\
&\times \int_0^{\infty} d R_{\rm CoM} R_{\rm CoM}^{2m+J+2} e^{-\frac{R_{\rm CoM}^2}{b^2}}
\left(\delta_{J0} \delta_{M0} + \frac{\alpha}{\sqrt{3}} R_{\rm CoM} \delta_{J1} \delta_{M0}\right) \\
&= \sqrt{\frac{\Gamma (N_r + 1)}{\Gamma (3/2) \Gamma (N_r + J + 3/2)}} \sum_{m=0}^{N_r} (-1)^{m} \binom{N_r + J + \frac{1}{2}}{N_r - m} \frac{1}{m!} \\
&\times \left[ \Gamma \left( \frac{2m+J+3}{2} \right) \delta_{J0} \delta_{M0}
+ \frac{\alpha b_{\rm CoM}}{\sqrt{3}} \Gamma \left( \frac{2m+J+4}{2} \right) \delta_{J1} \delta_{M0} \right] \, .
\end{split}
\end{equation}
%Knowing that
%\begin{equation}
%L_{N_r}^{J + \frac{1}{2}} \left( \frac{R_{\rm CoM}^{2}}{b^{2}} \right) = \sum_{m=0}^{N_r} (-)^{m} \binom{N_r + J + \frac{1}{2}}{N_r - m}
%\frac{1}{m!}\left(\frac{R_{\rm CoM}^{2}}{b^{2}}\right)^{m} \, ,
%\end{equation}
From the previous result we see that only the matrix elements with $M=0$ and $J=0,1$ are different from zero. They can finally be written as
\begin{align}
\braket{\Psi_{N 00}^{\rm CoM}|\Tilde{O}^{\rm CoM}|\Psi_{000}^{\rm CoM}} &= \sqrt{\frac{2\Gamma (N_r + 1) \Gamma (N_r + 3/2)}{\sqrt{\pi}}}
\sum_{m=0}^{N_r} (-1)^{m}  \frac{1}{\Gamma (N_r - m + 1) \Gamma (m+1)} \, , \\
\braket{\Psi_{N10}^{\rm CoM}|\Tilde{O}^{\rm CoM}|\Psi_{000}^{\rm CoM}} &= \alpha b_{\rm CoM} \sqrt{\frac{2 \Gamma (N_r + 1) \Gamma (N_r + 5/2)}{3 \sqrt{\pi}}}
\sum_{m=0}^{N_r} (-1)^{m} \frac{1}{\Gamma (N_r -m + 1) \Gamma (m+1)} \\
&= \alpha b_{\rm CoM} \sqrt{\frac{N_r + 3/2}{3}} \braket{\Psi_{N 00}^{\rm CoM}|\Tilde{O}^{\rm CoM}|\Psi_{000}^{\rm CoM}}
\, .
\end{align}
Now we want to evaluate the sum that occurs in the previous equations. To do this we consider the binomial coefficients
$(x+y)^n = \sum_{k=0}^n \binom{n}{k} x^{n-k} y^k$ with $x=1$ and $y=-1$, which give (for $n\geq 1$)
\begin{equation}
0 = \sum_{k=0}^n {(-1)}^k \binom{n}{k} = n! \sum_{k=0}^n {(-1)}^k \frac{1}{(n-k)! k!} \qquad \Rightarrow \qquad
\sum_{k=0}^n {(-1)}^k \frac{1}{(n-k)! k!} = 0 \, .
\end{equation}
We finally obtain the following results for the transition probabilities
\begin{align}
K_0^{\rm CoM} &= \left| \braket{\Psi_{000}^{\rm CoM}|\Tilde{O}^{\rm CoM}|\Psi_{000}^{\rm CoM}} \right|^2 = 1 \, , \\
K_1^{\rm CoM} &= \left| \braket{\Psi_{010}^{\rm CoM}|\Tilde{O}^{\rm CoM}|\Psi_{000}^{\rm CoM}} \right|^2 = \frac{\alpha^2 b_{\rm CoM}^2}{2} \, , \\
K_N^{\rm CoM} &= 0 \, , \quad N \geq 2 \, .
\end{align}

\end{widetext}

\bibliography{bibliography}

\begin{thebibliography}{38}%
\makeatletter
\providecommand \@ifxundefined [1]{%
 \@ifx{#1\undefined}
}%
\providecommand \@ifnum [1]{%
 \ifnum #1\expandafter \@firstoftwo
 \else \expandafter \@secondoftwo
 \fi
}%
\providecommand \@ifx [1]{%
 \ifx #1\expandafter \@firstoftwo
 \else \expandafter \@secondoftwo
 \fi
}%
\providecommand \natexlab [1]{#1}%
\providecommand \enquote  [1]{``#1''}%
\providecommand \bibnamefont  [1]{#1}%
\providecommand \bibfnamefont [1]{#1}%
\providecommand \citenamefont [1]{#1}%
\providecommand \href@noop [0]{\@secondoftwo}%
\providecommand \href [0]{\begingroup \@sanitize@url \@href}%
\providecommand \@href[1]{\@@startlink{#1}\@@href}%
\providecommand \@@href[1]{\endgroup#1\@@endlink}%
\providecommand \@sanitize@url [0]{\catcode `\\12\catcode `\$12\catcode
  `\&12\catcode `\#12\catcode `\^12\catcode `\_12\catcode `\%12\relax}%
\providecommand \@@startlink[1]{}%
\providecommand \@@endlink[0]{}%
\providecommand \url  [0]{\begingroup\@sanitize@url \@url }%
\providecommand \@url [1]{\endgroup\@href {#1}{\urlprefix }}%
\providecommand \urlprefix  [0]{URL }%
\providecommand \Eprint [0]{\href }%
\providecommand \doibase [0]{http://dx.doi.org/}%
\providecommand \selectlanguage [0]{\@gobble}%
\providecommand \bibinfo  [0]{\@secondoftwo}%
\providecommand \bibfield  [0]{\@secondoftwo}%
\providecommand \translation [1]{[#1]}%
\providecommand \BibitemOpen [0]{}%
\providecommand \bibitemStop [0]{}%
\providecommand \bibitemNoStop [0]{.\EOS\space}%
\providecommand \EOS [0]{\spacefactor3000\relax}%
\providecommand \BibitemShut  [1]{\csname bibitem#1\endcsname}%
\let\auto@bib@innerbib\@empty
%</preamble>
\bibitem [{\citenamefont {Jacobi}(1842)}]{Jacobi}%
  \BibitemOpen
  \bibfield  {author} {\bibinfo {author} {\bibfnamefont {C.~G.~J.}\
  \bibnamefont {Jacobi}},\ }\bibfield  {title} {\enquote {\bibinfo {title}
  {{Sur l'\'elimination des noeuds dans le probl\'eme des trois corps}},}\
  }\href@noop {} {\bibfield  {journal} {\bibinfo  {journal} {C. R. Acad. Sci.}\
  }\textbf {\bibinfo {volume} {15}},\ \bibinfo {pages} {236} (\bibinfo {year}
  {1842})}\BibitemShut {NoStop}%
\bibitem [{\citenamefont {Navr\'atil}\ \emph {et~al.}(2000)\citenamefont
  {Navr\'atil}, \citenamefont {Kamuntavi\ifmmode~\check{c}\else
  \v{c}\fi{}ius},\ and\ \citenamefont {Barrett}}]{Navratil2000}%
  \BibitemOpen
  \bibfield  {author} {\bibinfo {author} {\bibfnamefont {P.}~\bibnamefont
  {Navr\'atil}}, \bibinfo {author} {\bibfnamefont {G.~P.}\ \bibnamefont
  {Kamuntavi\ifmmode~\check{c}\else \v{c}\fi{}ius}}, \ and\ \bibinfo {author}
  {\bibfnamefont {B.~R.}\ \bibnamefont {Barrett}},\ }\bibfield  {title}
  {\enquote {\bibinfo {title} {Few-nucleon systems in a translationally
  invariant harmonic oscillator basis},}\ }\href {\doibase
  10.1103/PhysRevC.61.044001} {\bibfield  {journal} {\bibinfo  {journal} {Phys.
  Rev. C}\ }\textbf {\bibinfo {volume} {61}},\ \bibinfo {pages} {044001}
  (\bibinfo {year} {2000})}\BibitemShut {NoStop}%
\bibitem [{\citenamefont {Barnea}\ \emph {et~al.}(2001)\citenamefont {Barnea},
  \citenamefont {Leidemann},\ and\ \citenamefont {Orlandini}}]{barnea2001}%
  \BibitemOpen
  \bibfield  {author} {\bibinfo {author} {\bibfnamefont {N.}~\bibnamefont
  {Barnea}}, \bibinfo {author} {\bibfnamefont {W.}~\bibnamefont {Leidemann}}, \
  and\ \bibinfo {author} {\bibfnamefont {G.}~\bibnamefont {Orlandini}},\
  }\bibfield  {title} {\enquote {\bibinfo {title} {State-dependent effective
  interaction for the hyperspherical formalism with noncentral forces},}\
  }\href {\doibase http://dx.doi.org/10.1016/S0375-9474(01)00794-1} {\bibfield
  {journal} {\bibinfo  {journal} {Nuclear Physics A}\ }\textbf {\bibinfo
  {volume} {693}},\ \bibinfo {pages} {565 -- 578} (\bibinfo {year}
  {2001})}\BibitemShut {NoStop}%
\bibitem [{\citenamefont {Bacca}\ \emph
  {et~al.}(2009{\natexlab{a}})\citenamefont {Bacca}, \citenamefont {Schwenk},
  \citenamefont {Hagen},\ and\ \citenamefont {Papenbrock}}]{Bacca:2009yk}%
  \BibitemOpen
  \bibfield  {author} {\bibinfo {author} {\bibfnamefont {S.}~\bibnamefont
  {Bacca}}, \bibinfo {author} {\bibfnamefont {A.}~\bibnamefont {Schwenk}},
  \bibinfo {author} {\bibfnamefont {G.}~\bibnamefont {Hagen}}, \ and\ \bibinfo
  {author} {\bibfnamefont {T.}~\bibnamefont {Papenbrock}},\ }\bibfield  {title}
  {\enquote {\bibinfo {title} {{Helium halo nuclei from low-momentum
  interactions}},}\ }\href {\doibase 10.1140/epja/i2009-10815-5} {\bibfield
  {journal} {\bibinfo  {journal} {Eur. Phys. J. A}\ }\textbf {\bibinfo {volume}
  {42}},\ \bibinfo {pages} {553--558} (\bibinfo {year} {2009}{\natexlab{a}})},\
  \Eprint {http://arxiv.org/abs/0902.1696} {arXiv:0902.1696 [nucl-th]}
  \BibitemShut {NoStop}%
\bibitem [{\citenamefont {Bacca}(2013)}]{Bacca:2013ayo}%
  \BibitemOpen
  \bibfield  {author} {\bibinfo {author} {\bibfnamefont {S.}~\bibnamefont
  {Bacca}},\ }\bibfield  {title} {\enquote {\bibinfo {title} {{Neutron-rich
  Helium isotopes based on hyperspherical harmonics}},}\ }\href {\doibase
  10.22323/1.172.0119} {\bibfield  {journal} {\bibinfo  {journal} {PoS}\
  }\textbf {\bibinfo {volume} {CD12}},\ \bibinfo {pages} {119} (\bibinfo {year}
  {2013})},\ \Eprint {http://arxiv.org/abs/1302.2568} {arXiv:1302.2568
  [nucl-th]} \BibitemShut {NoStop}%
\bibitem [{\citenamefont {Slater}(1929)}]{Slater1929}%
  \BibitemOpen
  \bibfield  {author} {\bibinfo {author} {\bibfnamefont {J.~C.}\ \bibnamefont
  {Slater}},\ }\bibfield  {title} {\enquote {\bibinfo {title} {The theory of
  complex spectra},}\ }\href {\doibase 10.1103/PhysRev.34.1293} {\bibfield
  {journal} {\bibinfo  {journal} {Phys. Rev.}\ }\textbf {\bibinfo {volume}
  {34}},\ \bibinfo {pages} {1293--1322} (\bibinfo {year} {1929})}\BibitemShut
  {NoStop}%
\bibitem [{\citenamefont {de~Shalit}\ and\ \citenamefont
  {Talmi}(1963)}]{deShalit}%
  \BibitemOpen
  \bibfield  {author} {\bibinfo {author} {\bibfnamefont {A.}~\bibnamefont
  {de~Shalit}}\ and\ \bibinfo {author} {\bibfnamefont {I.}~\bibnamefont
  {Talmi}},\ }\href@noop {} {\emph {\bibinfo {title} {Nuclear shell theory}}}\
  (\bibinfo  {publisher} {Academic Press},\ \bibinfo {year} {1963})\BibitemShut
  {NoStop}%
\bibitem [{\citenamefont {Talmi}(1952)}]{Talmi}%
  \BibitemOpen
  \bibfield  {author} {\bibinfo {author} {\bibfnamefont {I.}~\bibnamefont
  {Talmi}},\ }\bibfield  {title} {\enquote {\bibinfo {title} {{Nuclear
  spectroscopy with harmonic oscillator wave-functions}},}\ }\href@noop {}
  {\bibfield  {journal} {\bibinfo  {journal} {Helv.~Phys. Acta}\ }\textbf
  {\bibinfo {volume} {25}},\ \bibinfo {pages} {185} (\bibinfo {year}
  {1952})}\BibitemShut {NoStop}%
\bibitem [{\citenamefont {Gartenhaus}\ and\ \citenamefont
  {Schwartz}(1957)}]{gartenhaus1957}%
  \BibitemOpen
  \bibfield  {author} {\bibinfo {author} {\bibfnamefont {S.}~\bibnamefont
  {Gartenhaus}}\ and\ \bibinfo {author} {\bibfnamefont {C.}~\bibnamefont
  {Schwartz}},\ }\bibfield  {title} {\enquote {\bibinfo {title} {Center-of-mass
  motion in many-particle systems},}\ }\href {\doibase 10.1103/PhysRev.108.482}
  {\bibfield  {journal} {\bibinfo  {journal} {Phys. Rev.}\ }\textbf {\bibinfo
  {volume} {108}},\ \bibinfo {pages} {482--490} (\bibinfo {year}
  {1957})}\BibitemShut {NoStop}%
\bibitem [{\citenamefont {Gloeckner}\ and\ \citenamefont
  {Lawson}(1974)}]{gloeckner1974}%
  \BibitemOpen
  \bibfield  {author} {\bibinfo {author} {\bibfnamefont {D.H.}\ \bibnamefont
  {Gloeckner}}\ and\ \bibinfo {author} {\bibfnamefont {R.D.}\ \bibnamefont
  {Lawson}},\ }\bibfield  {title} {\enquote {\bibinfo {title} {Spurious
  center-of-mass motion},}\ }\href {\doibase 10.1016/0370-2693(74)90390-6}
  {\bibfield  {journal} {\bibinfo  {journal} {Phys. Lett. B}\ }\textbf
  {\bibinfo {volume} {53}},\ \bibinfo {pages} {313 -- 318} (\bibinfo {year}
  {1974})}\BibitemShut {NoStop}%
\bibitem [{\citenamefont {McGrory}\ and\ \citenamefont
  {Wildenthal}(1975)}]{mcgrory1975}%
  \BibitemOpen
  \bibfield  {author} {\bibinfo {author} {\bibfnamefont {J.~B.}\ \bibnamefont
  {McGrory}}\ and\ \bibinfo {author} {\bibfnamefont {B.~H.}\ \bibnamefont
  {Wildenthal}},\ }\bibfield  {title} {\enquote {\bibinfo {title} {Further
  comment on spurious center-of-mass motion},}\ }\href {\doibase
  10.1016/0370-2693(75)90513-4} {\bibfield  {journal} {\bibinfo  {journal}
  {Phys. Lett. B}\ }\textbf {\bibinfo {volume} {60}},\ \bibinfo {pages} {5 --
  8} (\bibinfo {year} {1975})}\BibitemShut {NoStop}%
\bibitem [{\citenamefont {Roth}\ \emph {et~al.}(2009)\citenamefont {Roth},
  \citenamefont {Gour},\ and\ \citenamefont {Piecuch}}]{roth2009b}%
  \BibitemOpen
  \bibfield  {author} {\bibinfo {author} {\bibfnamefont {R.}~\bibnamefont
  {Roth}}, \bibinfo {author} {\bibfnamefont {J.~R.}\ \bibnamefont {Gour}}, \
  and\ \bibinfo {author} {\bibfnamefont {P.}~\bibnamefont {Piecuch}},\
  }\bibfield  {title} {\enquote {\bibinfo {title} {Center-of-mass problem in
  truncated configuration interaction and coupled-cluster calculations},}\
  }\href {\doibase 10.1016/j.physletb.2009.07.071} {\bibfield  {journal}
  {\bibinfo  {journal} {Phys. Lett. B}\ }\textbf {\bibinfo {volume} {679}},\
  \bibinfo {pages} {334 -- 339} (\bibinfo {year} {2009})}\BibitemShut {NoStop}%
\bibitem [{\citenamefont {Mihaila}\ and\ \citenamefont
  {Heisenberg}(1999)}]{mihaila1999}%
  \BibitemOpen
  \bibfield  {author} {\bibinfo {author} {\bibfnamefont {B.}~\bibnamefont
  {Mihaila}}\ and\ \bibinfo {author} {\bibfnamefont {J.~H.}\ \bibnamefont
  {Heisenberg}},\ }\bibfield  {title} {\enquote {\bibinfo {title}
  {Center-of-mass corrections reexamined: A many-body expansion approach},}\
  }\href {\doibase 10.1103/PhysRevC.60.054303} {\bibfield  {journal} {\bibinfo
  {journal} {Phys. Rev. C}\ }\textbf {\bibinfo {volume} {60}},\ \bibinfo
  {pages} {054303} (\bibinfo {year} {1999})}\BibitemShut {NoStop}%
\bibitem [{\citenamefont {Hagen}\ \emph {et~al.}(2009)\citenamefont {Hagen},
  \citenamefont {Papenbrock},\ and\ \citenamefont {Dean}}]{hagen2009a}%
  \BibitemOpen
  \bibfield  {author} {\bibinfo {author} {\bibfnamefont {G.}~\bibnamefont
  {Hagen}}, \bibinfo {author} {\bibfnamefont {T.}~\bibnamefont {Papenbrock}}, \
  and\ \bibinfo {author} {\bibfnamefont {D.~J.}\ \bibnamefont {Dean}},\
  }\bibfield  {title} {\enquote {\bibinfo {title} {Solution of the
  center-of-mass problem in nuclear structure calculations},}\ }\href {\doibase
  10.1103/PhysRevLett.103.062503} {\bibfield  {journal} {\bibinfo  {journal}
  {Phys. Rev. Lett.}\ }\textbf {\bibinfo {volume} {103}},\ \bibinfo {pages}
  {062503} (\bibinfo {year} {2009})}\BibitemShut {NoStop}%
\bibitem [{\citenamefont {Baker}\ \emph {et~al.}(2020)\citenamefont {Baker},
  \citenamefont {Launey}, \citenamefont {Bacca}, \citenamefont {Dinur},\ and\
  \citenamefont {Dytrych}}]{Baker2020}%
  \BibitemOpen
  \bibfield  {author} {\bibinfo {author} {\bibfnamefont {R.~B.}\ \bibnamefont
  {Baker}}, \bibinfo {author} {\bibfnamefont {K.~D.}\ \bibnamefont {Launey}},
  \bibinfo {author} {\bibfnamefont {S.}~\bibnamefont {Bacca}}, \bibinfo
  {author} {\bibfnamefont {N.~Nevo}\ \bibnamefont {Dinur}}, \ and\ \bibinfo
  {author} {\bibfnamefont {T.}~\bibnamefont {Dytrych}},\ }\bibfield  {title}
  {\enquote {\bibinfo {title} {Benchmark calculations of electromagnetic sum
  rules with a symmetry-adapted basis and hyperspherical harmonics},}\ }\href
  {\doibase 10.1103/PhysRevC.102.014320} {\bibfield  {journal} {\bibinfo
  {journal} {Phys. Rev. C}\ }\textbf {\bibinfo {volume} {102}},\ \bibinfo
  {pages} {014320} (\bibinfo {year} {2020})}\BibitemShut {NoStop}%
\bibitem [{\citenamefont {Bacca}\ \emph {et~al.}(2007)\citenamefont {Bacca},
  \citenamefont {Arenh\"ovel}, \citenamefont {Barnea}, \citenamefont
  {Leidemann},\ and\ \citenamefont {Orlandini}}]{th4He}%
  \BibitemOpen
  \bibfield  {author} {\bibinfo {author} {\bibfnamefont {S.}~\bibnamefont
  {Bacca}}, \bibinfo {author} {\bibfnamefont {H.}~\bibnamefont {Arenh\"ovel}},
  \bibinfo {author} {\bibfnamefont {N.}~\bibnamefont {Barnea}}, \bibinfo
  {author} {\bibfnamefont {W.}~\bibnamefont {Leidemann}}, \ and\ \bibinfo
  {author} {\bibfnamefont {G.}~\bibnamefont {Orlandini}},\ }\bibfield  {title}
  {\enquote {\bibinfo {title} {Inclusive electron scattering off
  $^{4}\mathrm{He}$},}\ }\href {\doibase 10.1103/PhysRevC.76.014003} {\bibfield
   {journal} {\bibinfo  {journal} {Phys. Rev. C}\ }\textbf {\bibinfo {volume}
  {76}},\ \bibinfo {pages} {014003} (\bibinfo {year} {2007})}\BibitemShut
  {NoStop}%
\bibitem [{\citenamefont {Acharya}\ and\ \citenamefont {Bacca}(2020)}]{bijaya}%
  \BibitemOpen
  \bibfield  {author} {\bibinfo {author} {\bibfnamefont {B.}~\bibnamefont
  {Acharya}}\ and\ \bibinfo {author} {\bibfnamefont {S.}~\bibnamefont
  {Bacca}},\ }\bibfield  {title} {\enquote {\bibinfo {title} {Neutrino-deuteron
  scattering: Uncertainty quantification and new ${L}_{1,A}$ constraints},}\
  }\href {\doibase 10.1103/PhysRevC.101.015505} {\bibfield  {journal} {\bibinfo
   {journal} {Phys. Rev. C}\ }\textbf {\bibinfo {volume} {101}},\ \bibinfo
  {pages} {015505} (\bibinfo {year} {2020})}\BibitemShut {NoStop}%
\bibitem [{\citenamefont {Bacca}\ and\ \citenamefont
  {Pastore}(2014)}]{Bacca_2014}%
  \BibitemOpen
  \bibfield  {author} {\bibinfo {author} {\bibfnamefont {S.}~\bibnamefont
  {Bacca}}\ and\ \bibinfo {author} {\bibfnamefont {S.}~\bibnamefont
  {Pastore}},\ }\bibfield  {title} {\enquote {\bibinfo {title} {Electromagnetic
  reactions on light nuclei},}\ }\href {\doibase
  10.1088/0954-3899/41/12/123002} {\bibfield  {journal} {\bibinfo  {journal}
  {J.~Phys.~G: Nucl.~Part.~Phys.}\ }\textbf {\bibinfo {volume} {41}},\ \bibinfo
  {pages} {123002} (\bibinfo {year} {2014})}\BibitemShut {NoStop}%
\bibitem [{\citenamefont {Gazit}\ \emph {et~al.}(2006)\citenamefont {Gazit},
  \citenamefont {Bacca}, \citenamefont {Barnea}, \citenamefont {Leidemann},\
  and\ \citenamefont {Orlandini}}]{gazit2006}%
  \BibitemOpen
  \bibfield  {author} {\bibinfo {author} {\bibfnamefont {D.}~\bibnamefont
  {Gazit}}, \bibinfo {author} {\bibfnamefont {S.}~\bibnamefont {Bacca}},
  \bibinfo {author} {\bibfnamefont {N.}~\bibnamefont {Barnea}}, \bibinfo
  {author} {\bibfnamefont {W.}~\bibnamefont {Leidemann}}, \ and\ \bibinfo
  {author} {\bibfnamefont {G.}~\bibnamefont {Orlandini}},\ }\bibfield  {title}
  {\enquote {\bibinfo {title} {Photoabsorption on $^{4}\mathrm{He}$ with a
  realistic nuclear force},}\ }\href {\doibase 10.1103/PhysRevLett.96.112301}
  {\bibfield  {journal} {\bibinfo  {journal} {Phys. Rev. Lett.}\ }\textbf
  {\bibinfo {volume} {96}},\ \bibinfo {pages} {112301} (\bibinfo {year}
  {2006})}\BibitemShut {NoStop}%
\bibitem [{\citenamefont {Bacca}\ \emph
  {et~al.}(2009{\natexlab{b}})\citenamefont {Bacca}, \citenamefont {Barnea},
  \citenamefont {Leidemann},\ and\ \citenamefont {Orlandini}}]{BaccaPRL2009}%
  \BibitemOpen
  \bibfield  {author} {\bibinfo {author} {\bibfnamefont {S.}~\bibnamefont
  {Bacca}}, \bibinfo {author} {\bibfnamefont {N.}~\bibnamefont {Barnea}},
  \bibinfo {author} {\bibfnamefont {W.}~\bibnamefont {Leidemann}}, \ and\
  \bibinfo {author} {\bibfnamefont {G.}~\bibnamefont {Orlandini}},\ }\bibfield
  {title} {\enquote {\bibinfo {title} {Role of the final-state interaction and
  three-body force on the longitudinal response function of
  $^{4}\mathrm{He}$},}\ }\href {\doibase 10.1103/PhysRevLett.102.162501}
  {\bibfield  {journal} {\bibinfo  {journal} {Phys. Rev. Lett.}\ }\textbf
  {\bibinfo {volume} {102}},\ \bibinfo {pages} {162501} (\bibinfo {year}
  {2009}{\natexlab{b}})}\BibitemShut {NoStop}%
\bibitem [{\citenamefont {Bacca}\ \emph
  {et~al.}(2013{\natexlab{a}})\citenamefont {Bacca}, \citenamefont {Barnea},
  \citenamefont {Leidemann},\ and\ \citenamefont {Orlandini}}]{BaccaMono}%
  \BibitemOpen
  \bibfield  {author} {\bibinfo {author} {\bibfnamefont {S.}~\bibnamefont
  {Bacca}}, \bibinfo {author} {\bibfnamefont {N.}~\bibnamefont {Barnea}},
  \bibinfo {author} {\bibfnamefont {W.}~\bibnamefont {Leidemann}}, \ and\
  \bibinfo {author} {\bibfnamefont {G.}~\bibnamefont {Orlandini}},\ }\bibfield
  {title} {\enquote {\bibinfo {title} {Isoscalar monopole resonance of the
  alpha particle: A prism to nuclear hamiltonians},}\ }\href {\doibase
  10.1103/PhysRevLett.110.042503} {\bibfield  {journal} {\bibinfo  {journal}
  {Phys. Rev. Lett.}\ }\textbf {\bibinfo {volume} {110}},\ \bibinfo {pages}
  {042503} (\bibinfo {year} {2013}{\natexlab{a}})}\BibitemShut {NoStop}%
\bibitem [{\citenamefont {Barrett}\ \emph {et~al.}(2013)\citenamefont
  {Barrett}, \citenamefont {Navratil},\ and\ \citenamefont
  {Vary}}]{Barrett:2013nh}%
  \BibitemOpen
  \bibfield  {author} {\bibinfo {author} {\bibfnamefont {B.~R.}\ \bibnamefont
  {Barrett}}, \bibinfo {author} {\bibfnamefont {P.}~\bibnamefont {Navratil}}, \
  and\ \bibinfo {author} {\bibfnamefont {J.~P.}\ \bibnamefont {Vary}},\
  }\bibfield  {title} {\enquote {\bibinfo {title} {{Ab initio no core shell
  model}},}\ }\href {\doibase 10.1016/j.ppnp.2012.10.003} {\bibfield  {journal}
  {\bibinfo  {journal} {Prog. Part. Nucl. Phys.}\ }\textbf {\bibinfo {volume}
  {69}},\ \bibinfo {pages} {131--181} (\bibinfo {year} {2013})}\BibitemShut
  {NoStop}%
\bibitem [{\citenamefont {Navratil}\ \emph {et~al.}(2009)\citenamefont
  {Navratil}, \citenamefont {Quaglioni}, \citenamefont {Stetcu},\ and\
  \citenamefont {Barrett}}]{Navratil:2009ut}%
  \BibitemOpen
  \bibfield  {author} {\bibinfo {author} {\bibfnamefont {P.}~\bibnamefont
  {Navratil}}, \bibinfo {author} {\bibfnamefont {S.}~\bibnamefont {Quaglioni}},
  \bibinfo {author} {\bibfnamefont {I.}~\bibnamefont {Stetcu}}, \ and\ \bibinfo
  {author} {\bibfnamefont {B.~R.}\ \bibnamefont {Barrett}},\ }\bibfield
  {title} {\enquote {\bibinfo {title} {{Recent developments in no-core
  shell-model calculations}},}\ }\href {\doibase 10.1088/0954-3899/36/8/083101}
  {\bibfield  {journal} {\bibinfo  {journal} {J. Phys. G}\ }\textbf {\bibinfo
  {volume} {36}},\ \bibinfo {pages} {083101} (\bibinfo {year} {2009})},\
  \Eprint {http://arxiv.org/abs/0904.0463} {arXiv:0904.0463 [nucl-th]}
  \BibitemShut {NoStop}%
\bibitem [{\citenamefont {Varshalovich}\ \emph {et~al.}(1988)\citenamefont
  {Varshalovich}, \citenamefont {Moskalev},\ and\ \citenamefont
  {Khersonskii}}]{Varshalovich}%
  \BibitemOpen
  \bibfield  {author} {\bibinfo {author} {\bibfnamefont {D.~A.}\ \bibnamefont
  {Varshalovich}}, \bibinfo {author} {\bibfnamefont {A.~N.}\ \bibnamefont
  {Moskalev}}, \ and\ \bibinfo {author} {\bibfnamefont {V.~K.}\ \bibnamefont
  {Khersonskii}},\ }\href@noop {} {\emph {\bibinfo {title} {Quantum Theory of
  Angular Momentum}}}\ (\bibinfo  {publisher} {World Scientific},\ \bibinfo
  {address} {Singapore},\ \bibinfo {year} {1988})\BibitemShut {NoStop}%
\bibitem [{\citenamefont {Edmonds}(1996)}]{edmonds1996angular}%
  \BibitemOpen
  \bibfield  {author} {\bibinfo {author} {\bibfnamefont {A.R.}\ \bibnamefont
  {Edmonds}},\ }\href {https://books.google.de/books?id=0BSOg0oHhZ0C} {\emph
  {\bibinfo {title} {Angular Momentum in Quantum Mechanics}}},\ Investigations
  in Physics Series\ (\bibinfo  {publisher} {Princeton University Press},\
  \bibinfo {year} {1996})\BibitemShut {NoStop}%
\bibitem [{\citenamefont {Miorelli}\ \emph {et~al.}(2016)\citenamefont
  {Miorelli}, \citenamefont {Bacca}, \citenamefont {Barnea}, \citenamefont
  {Hagen}, \citenamefont {Jansen}, \citenamefont {Orlandini},\ and\
  \citenamefont {Papenbrock}}]{miorelli2016}%
  \BibitemOpen
  \bibfield  {author} {\bibinfo {author} {\bibfnamefont {M.}~\bibnamefont
  {Miorelli}}, \bibinfo {author} {\bibfnamefont {S.}~\bibnamefont {Bacca}},
  \bibinfo {author} {\bibfnamefont {N.}~\bibnamefont {Barnea}}, \bibinfo
  {author} {\bibfnamefont {G.}~\bibnamefont {Hagen}}, \bibinfo {author}
  {\bibfnamefont {G.~R.}\ \bibnamefont {Jansen}}, \bibinfo {author}
  {\bibfnamefont {G.}~\bibnamefont {Orlandini}}, \ and\ \bibinfo {author}
  {\bibfnamefont {T.}~\bibnamefont {Papenbrock}},\ }\bibfield  {title}
  {\enquote {\bibinfo {title} {Electric dipole polarizability from first
  principles calculations},}\ }\href {\doibase 10.1103/PhysRevC.94.034317}
  {\bibfield  {journal} {\bibinfo  {journal} {Phys. Rev. C}\ }\textbf {\bibinfo
  {volume} {94}},\ \bibinfo {pages} {034317} (\bibinfo {year}
  {2016})}\BibitemShut {NoStop}%
\bibitem [{\citenamefont {Miorelli}\ \emph {et~al.}(2018)\citenamefont
  {Miorelli}, \citenamefont {Bacca}, \citenamefont {Hagen},\ and\ \citenamefont
  {Papenbrock}}]{miorelli2018}%
  \BibitemOpen
  \bibfield  {author} {\bibinfo {author} {\bibfnamefont {M.}~\bibnamefont
  {Miorelli}}, \bibinfo {author} {\bibfnamefont {S.}~\bibnamefont {Bacca}},
  \bibinfo {author} {\bibfnamefont {G.}~\bibnamefont {Hagen}}, \ and\ \bibinfo
  {author} {\bibfnamefont {T.}~\bibnamefont {Papenbrock}},\ }\bibfield  {title}
  {\enquote {\bibinfo {title} {Computing the dipole polarizability of
  $^{48}\mathrm{Ca}$ with increased precision},}\ }\href {\doibase
  10.1103/PhysRevC.98.014324} {\bibfield  {journal} {\bibinfo  {journal} {Phys.
  Rev. C}\ }\textbf {\bibinfo {volume} {98}},\ \bibinfo {pages} {014324}
  (\bibinfo {year} {2018})}\BibitemShut {NoStop}%
\bibitem [{\citenamefont {Suhonen}(2007)}]{-Suhonen-J}%
  \BibitemOpen
  \bibfield  {author} {\bibinfo {author} {\bibfnamefont {J.}~\bibnamefont
  {Suhonen}},\ }\href {https://books.google.de/books?id=Cye7s8LH-IkC} {\emph
  {\bibinfo {title} {From Nucleons to Nucleus}}},\ Theoretical and Mathematical
  Physics\ (\bibinfo  {publisher} {Springer Berlin Heidelberg},\ \bibinfo
  {year} {2007})\BibitemShut {NoStop}%
\bibitem [{\citenamefont {Xu}(2016)}]{tianrui}%
  \BibitemOpen
  \bibfield  {author} {\bibinfo {author} {\bibfnamefont {T.}~\bibnamefont
  {Xu}},\ }\href@noop {} {\enquote {\bibinfo {title} {Electromagnetic sum rules
  in light nuclei},}\ }\bibinfo {howpublished} {Bachelor Thesis} (\bibinfo
  {year} {2016}),\ \bibinfo {note} {at the University of British
  Columbia}\BibitemShut {NoStop}%
\bibitem [{\citenamefont {Weinberger}(2017)}]{israel}%
  \BibitemOpen
  \bibfield  {author} {\bibinfo {author} {\bibfnamefont {I.}~\bibnamefont
  {Weinberger}},\ }\href@noop {} {\enquote {\bibinfo {title} {Separaring center
  of mass effects from calculated many-body reaction cross section},}\
  }\bibinfo {howpublished} {Master Thesis} (\bibinfo {year} {2017}),\ \bibinfo
  {note} {at the Hebrew University}\BibitemShut {NoStop}%
\bibitem [{\citenamefont {Sobczyk}\ \emph {et~al.}(2020)\citenamefont
  {Sobczyk}, \citenamefont {Acharya}, \citenamefont {Bacca},\ and\
  \citenamefont {Hagen}}]{Sobczyk2020}%
  \BibitemOpen
  \bibfield  {author} {\bibinfo {author} {\bibfnamefont {J.~E.}\ \bibnamefont
  {Sobczyk}}, \bibinfo {author} {\bibfnamefont {B.}~\bibnamefont {Acharya}},
  \bibinfo {author} {\bibfnamefont {S.}~\bibnamefont {Bacca}}, \ and\ \bibinfo
  {author} {\bibfnamefont {G.}~\bibnamefont {Hagen}},\ }\bibfield  {title}
  {\enquote {\bibinfo {title} {Coulomb sum rule for $^{4}\mathrm{He}$ and
  $^{16}\mathrm{O}$ from coupled-cluster theory},}\ }\href {\doibase
  10.1103/PhysRevC.102.064312} {\bibfield  {journal} {\bibinfo  {journal}
  {Phys. Rev. C}\ }\textbf {\bibinfo {volume} {102}},\ \bibinfo {pages}
  {064312} (\bibinfo {year} {2020})}\BibitemShut {NoStop}%
\bibitem [{\citenamefont {Bacca}\ \emph
  {et~al.}(2013{\natexlab{b}})\citenamefont {Bacca}, \citenamefont {Barnea},
  \citenamefont {Hagen}, \citenamefont {Orlandini},\ and\ \citenamefont
  {Papenbrock}}]{bacca2013}%
  \BibitemOpen
  \bibfield  {author} {\bibinfo {author} {\bibfnamefont {S.}~\bibnamefont
  {Bacca}}, \bibinfo {author} {\bibfnamefont {N.}~\bibnamefont {Barnea}},
  \bibinfo {author} {\bibfnamefont {G.}~\bibnamefont {Hagen}}, \bibinfo
  {author} {\bibfnamefont {G.}~\bibnamefont {Orlandini}}, \ and\ \bibinfo
  {author} {\bibfnamefont {T.}~\bibnamefont {Papenbrock}},\ }\bibfield  {title}
  {\enquote {\bibinfo {title} {First principles description of the giant dipole
  resonance in $^{16}\mathbf{O}$},}\ }\href {\doibase
  10.1103/PhysRevLett.111.122502} {\bibfield  {journal} {\bibinfo  {journal}
  {Phys. Rev. Lett.}\ }\textbf {\bibinfo {volume} {111}},\ \bibinfo {pages}
  {122502} (\bibinfo {year} {2013}{\natexlab{b}})}\BibitemShut {NoStop}%
\bibitem [{\citenamefont {Bacca}\ \emph {et~al.}(2014)\citenamefont {Bacca},
  \citenamefont {Barnea}, \citenamefont {Hagen}, \citenamefont {Miorelli},
  \citenamefont {Orlandini},\ and\ \citenamefont {Papenbrock}}]{bacca2014}%
  \BibitemOpen
  \bibfield  {author} {\bibinfo {author} {\bibfnamefont {S.}~\bibnamefont
  {Bacca}}, \bibinfo {author} {\bibfnamefont {N.}~\bibnamefont {Barnea}},
  \bibinfo {author} {\bibfnamefont {G.}~\bibnamefont {Hagen}}, \bibinfo
  {author} {\bibfnamefont {M.}~\bibnamefont {Miorelli}}, \bibinfo {author}
  {\bibfnamefont {G.}~\bibnamefont {Orlandini}}, \ and\ \bibinfo {author}
  {\bibfnamefont {T.}~\bibnamefont {Papenbrock}},\ }\bibfield  {title}
  {\enquote {\bibinfo {title} {Giant and pigmy dipole resonances in
  $^{4}\mathrm{He}$, $^{16,22}\mathrm{O}$, and $^{40}\mathrm{Ca}$ from chiral
  nucleon-nucleon interactions},}\ }\href {\doibase 10.1103/PhysRevC.90.064619}
  {\bibfield  {journal} {\bibinfo  {journal} {Phys. Rev. C}\ }\textbf {\bibinfo
  {volume} {90}},\ \bibinfo {pages} {064619} (\bibinfo {year}
  {2014})}\BibitemShut {NoStop}%
\bibitem [{\citenamefont {Entem}\ and\ \citenamefont
  {Machleidt}(2003)}]{entem2003}%
  \BibitemOpen
  \bibfield  {author} {\bibinfo {author} {\bibfnamefont {D.~R.}\ \bibnamefont
  {Entem}}\ and\ \bibinfo {author} {\bibfnamefont {R.}~\bibnamefont
  {Machleidt}},\ }\bibfield  {title} {\enquote {\bibinfo {title} {Accurate
  charge-dependent nucleon-nucleon potential at fourth order of chiral
  perturbation theory},}\ }\href {\doibase 10.1103/PhysRevC.68.041001}
  {\bibfield  {journal} {\bibinfo  {journal} {Phys. Rev. C}\ }\textbf {\bibinfo
  {volume} {68}},\ \bibinfo {pages} {041001} (\bibinfo {year}
  {2003})}\BibitemShut {NoStop}%
\bibitem [{\citenamefont {Audi}\ \emph {et~al.}(2003)\citenamefont {Audi},
  \citenamefont {Bersillon}, \citenamefont {Blachot},\ and\ \citenamefont
  {Wapstra}}]{AUDI20033}%
  \BibitemOpen
  \bibfield  {author} {\bibinfo {author} {\bibfnamefont {G.}~\bibnamefont
  {Audi}}, \bibinfo {author} {\bibfnamefont {O.}~\bibnamefont {Bersillon}},
  \bibinfo {author} {\bibfnamefont {J.}~\bibnamefont {Blachot}}, \ and\
  \bibinfo {author} {\bibfnamefont {A.H.}\ \bibnamefont {Wapstra}},\ }\bibfield
   {title} {\enquote {\bibinfo {title} {The nubase evaluation of nuclear and
  decay properties},}\ }\href {\doibase
  https://doi.org/10.1016/j.nuclphysa.2003.11.001} {\bibfield  {journal}
  {\bibinfo  {journal} {Nuclear Physics A}\ }\textbf {\bibinfo {volume}
  {729}},\ \bibinfo {pages} {3 -- 128} (\bibinfo {year} {2003})},\ \bibinfo
  {note} {the 2003 NUBASE and Atomic Mass Evaluations}\BibitemShut {NoStop}%
\bibitem [{\citenamefont {Layh}(2020)}]{dennis}%
  \BibitemOpen
  \bibfield  {author} {\bibinfo {author} {\bibfnamefont {D.}~\bibnamefont
  {Layh}},\ }\href@noop {} {\enquote {\bibinfo {title} {Electromagnetic sum
  rules for the nuclear two-body system in the intrinsic frame},}\ }\bibinfo
  {howpublished} {Bachelor Thesis} (\bibinfo {year} {2020}),\ \bibinfo {note}
  {at Johannes Gutenberg Universit\"at Mainz}\BibitemShut {NoStop}%
\bibitem [{\citenamefont {Akdogan}(2021)}]{kerem}%
  \BibitemOpen
  \bibfield  {author} {\bibinfo {author} {\bibfnamefont {K.}~\bibnamefont
  {Akdogan}},\ }\href@noop {} {\enquote {\bibinfo {title} {Electromagnetic sum
  rules for the nuclear two-body system in the laboratory frame},}\ }\bibinfo
  {howpublished} {Bachelor Thesis} (\bibinfo {year} {2021}),\ \bibinfo {note}
  {at Johannes Gutenberg Universit\"at Mainz}\BibitemShut {NoStop}%
\bibitem [{\citenamefont {Abramowitz}\ and\ \citenamefont
  {Stegun}(1964)}]{Abramowitz}%
  \BibitemOpen
  \bibfield  {author} {\bibinfo {author} {\bibfnamefont {M.}~\bibnamefont
  {Abramowitz}}\ and\ \bibinfo {author} {\bibfnamefont {I.~A.}\ \bibnamefont
  {Stegun}},\ }\href@noop {} {\emph {\bibinfo {title} {Handbook of Mathematical
  Functions with Formulas, Graphs, and Mathematical Tables}}},\ \bibinfo
  {edition} {ninth dover printing, tenth gpo printing}\ ed.\ (\bibinfo
  {publisher} {Dover},\ \bibinfo {address} {New York},\ \bibinfo {year}
  {1964})\BibitemShut {NoStop}%
\end{thebibliography}%

\end{document}